\documentstyle[aps,prb,multicol]{revtex}
\begin{document}
\title{Electron-phonon interactions and related physical properties of metals from
linear-response theory}
\author{S. Y. Savrasov}
\address{Max-Planck-Institut f\"ur Festk\"orperforschung, Heisenbergstr. 1, 70569
Stuttgart, Germany.}
\author{D. Y. Savrasov}
\address{P. N. Lebedev Physical Institute, Leninskii pr.53, 117924 Moscow, Russia}
\date{\today }
\maketitle

\begin{abstract}
Spectral distribution functions of electron-phonon interaction $\alpha
^2F(\omega )$ obtained by {\it ab initio }linear--response calculations are
used to describe various superconducting and transport properties in a
number of elemental metals such as $Al,Cu,Mo,Nb,Pb,Pd,Ta,$ and $V.$ Their
lattice dynamics and self--consistently screened electron--phonon coupling
are evaluated within local density functional theory and using
linear--muffin-tin--orbital basis set. We compare our theoretical $\alpha
^2F(\omega )$ with those deduced from the tunneling measurements and find a
close agreement between them. Temperature dependent electrical and thermal
resistivities as well as transport constants $\lambda _{tr}$ also agree well
with the experimental data. The values of $\lambda _{tr}$ are close to the
electron-phonon coupling parameter $\lambda .$ For the later a very good
agrement with specific-heat measurements was found without any paramagnon
contribution, except in $Pd$. We conclude that our method provides the
description of electron-phonon interactions in tested materials with the
accuracy $10\%$.
\end{abstract}


%\twocolumn
%TCIMACRO{\TeXButton{twocolumn}{\begin{multicols}{2}}}
%BeginExpansion
\begin{multicols}{2}
%EndExpansion

\smallskip\ 

{\bf I. INTRODUCTION.}

\smallskip\ 

Electron-phonon interaction (EPI) in metals is a subject of intensive
theoretical and experimental investigations. The interest to this problem
arises from a variety of physical phenomena such as electrical and thermal
resistivities, renormalization of the electronic specific heat (electronic
mass enhancement), and, of course, superconductivity, for quantitative
understanding of those a proper description of the EPI is required\cite
{Grimvall}. Moreover, the reliable estimation of the EPI parameters in
particular case of high-temperature superconductivity may be decisive for
recognizing the nature of this phenomenon. Unfortunately, even for some
transition metals we have controversial experimental and theoretical data
related to the estimation of the coupling constant $\lambda $. Analysis is
complicated by possible parallel processes of spin-fluctuations, for
example, in the problem\cite{Parks} of renormalization of specific heat and $%
T_c$ or proximity effects in the tunneling data\cite{Wolf}. To extract the
quantities under the interest one has to use some theoretical calculations
and models. In this situation fully {\it ab initio} calculations of the
one-electron spectra and phonon dispersions based on density functional
theory\cite{DFT} (DFT) are most preferable. Applicability of the popular
local density approximation\cite{LDA} (LDA) for the functional and treatment
of the one-electron band structures as spectra of low-energy electronic
excitations were checked many times and there exist a reach theoretical and
experimental experience\cite{GAS}. This allows us to conclude that even
having been formally ground--state theory, DFT is a good starting point for
investigating the electron-phonon interaction.

Many previous attempts to compute EPI, in particular, for transition metals,
have focused on calculating merely the electronic contribution to this
quantity\cite{Pickett}, while the phonon frequencies $\omega _{{\bf q}\nu }$
and the eigenvectors ${\bf \eta }_{{\bf q}\nu }$ were usually taken from
inelastic neutron-scattering data. There, the self-consistent adjustment of
the one-electron potential to the phonon distortion was replaced by either
rigid-ion approximation (RIA)\cite{RIA} or most popular rigid-muffin-tin
approximation\cite{RMTA} (RMTA). For isotropic metals, having a large
density of states at the Fermi energy, the RMTA works well in many cases,
since the efficient electronic screening limits the change of the potential
in the immediate vicinity of the displaced atom. However, there are known
some problems of the RMTA in transition metals. For example, anisotropy of
the mass enhancement factor was not reproduced by the RMTA in $Nb$\cite
{Anisotropy}.

Accurate $\omega _{{\bf q}\nu }$ and ${\bf \eta }_{{\bf q}\nu }$ as well as
self--consistently screened electron--phonon interaction can be calculated
within the total--energy frozen--phonon approach using the supercells\cite
{Devreese,Dacorogna,Cohen,Liecht}. However, there is a serious drawback of
this method. Sufficiently large number of phonon wave vectors ${\bf q\,}$%
must be sampled in the Brillouin zone to get a good estimate of the average
coupling strength $\lambda $. A separate frozen--phonon calculation is
required for each ${\bf q}$ and for each studied mode. For small phonon wave
vectors{\bf \ }this requires very large supercells. With the crude sampling
allowed by the limited size of the supercell, the accuracy of ${\bf q}$
integrated quantities like $\lambda $ is usually low.

Another technique which can be employed for calculating the self--consistent
change in the potential is the perturbative approach\cite{Devreese}
applicable for any ${\bf q}$ . Key quantity of this method is an
independent--particle polarizability function. After applying the
first--order perturbation theory and expanding the first--order changes in
the one--electron wave functions over the basis of unperturbed Bloch states,
the polarizability is expressed via the double sum over occupied and
unoccupied states. Winter\cite{Winter} has successfully applied this method
to calculate $\lambda $ in $Al$. Unfortunately, the perturbative approach
has several drawbacks. First, the slowly convergent sum over the excited
states requires their preliminary calculation by diagonalizing the
unperturbed hamiltonian matrix of very large dimension. Second, the
self--consistency in this method is done by inverting the dielectric matrix
of the crystal which is a relatively time-consuming problem.

To date, the most efficient technique developed to calculate the lattice
dynamics is the solid--state generalization\cite{Baroni} of the Sternheimer
method\cite{Sternheimer}. This method is not limited to ${\bf q}$'s
commensurate with the lattice as the frozen-phonon approach and it does not
require the knowledge of all unperturbed states as the perturbative
approach. The latter advantage can be achieved by constructing a rapidly
convergent basis set for representing the first-order corrections to the
wave functions other than the basis of unperturbed states. This is important
because the first-order corrections as well as the unperturbed wave
functions oscillate in the core region. In broad-band semiconductors and
insulators this problem can be circumvented by using the pseudopotential
method and most calculations of phonon dispersions performed so far use
plane-wave basis sets\cite{APPL}. Unfortunately with decreasing band width
the plane-wave expansion of the pseudo-wave-functions converges more slowly
and it becomes less advantageous to use pseudopotentials.

Recently an all-electron formulation of this generalized Sternheimer
approach has been given\cite{LR,UFN,LRPRB}. The first-order corrections are
represented in terms of muffin-tin (MT) basis set such as
linear-muffin-tin-orbitals\cite{OKA} (LMTO) which greatly facilitates the
treatment of localized valence wave functions. For the first time, the
method was shown to produce accurate phonon dispersions in transition metals
and transition-metal compounds\cite{LR,UFN,LRPRB}. In this paper we present
details of generalizing this method to compute the wave--vector--dependent
electron-phonon coupling. (A brief report of this work has appeared already%
\cite{EPI}). We evaluate the spectral distribution functions $\alpha
^2F(\omega )$ of the EPI from the phonon linewidths $\gamma _{{\bf q}\nu }$
according to the approach developed by Allen\cite{Allen} in the
superconductivity theory\cite{Eliashberg}. The electron--phonon matrix
elements are calculated in the LMTO representation. Due to incompleteness of
the basis sets in band--structure calculations the corrections to these
matrix elements are shown to exist and explicitly taken into account. The 
incomplete-basis-set (IBS) corrections appear here in the same manner as in
the calculation of the dynamical matrix within the linear--response theory%
\cite{LR} or when calculating the forces within the total--energy
frozen--phonon approach in terms of the LMTO\ method\cite{Forces}.

We apply the developed scheme to compute electron--phonon coupling for a
large number of elemental metals. We also present calculations of their
phonon dispersion curves. The results of computed transport properties such
as temperature-dependent phonon-limited electrical resistivities and thermal
conductivities obtained as low-order variational solutions of the Boltzman
equation are also given. The method of calculating the transport properties
is analogous to that used in the supeconductivity theory and is based\cite
{AllenTR} on calculating the transport spectral functions $\alpha
_{tr}^2F(\omega )$. All the results presented in this paper are completely 
{\it ab initio} and no adjustable parameters have been used in the
calculations.

The rest of the paper is organized as follows. In section II we derive the
formulae for calculating the electron-phonon matrix elements and briefly
review the method of finding superconducting and transport properties.
Section III presents the results of the calculations for phonon dispersions,
electron-phonon interactions and related properties for a number of
elemental metals such as $Al,Cu,Mo,Nb,Pb,Pd,Ta,$ and $V.$ Section IV
concludes the paper.

\smallskip\ 

{\bf II. METHOD.}

\smallskip\ 

The central problem in calculating the electron-phonon interaction is the
evaluation of changes in the electronic Hamiltonian caused by atomic
displacements. This generally requires the knowledge of the full low--energy
excitation spectrum of the metal: the quasiparticle energies and the phonon
frequencies$.$ The calculations of vibrational properties are, in principle,
within the scope of the density--functional based methods. Finding the
quasiparticle excitation spectra is, on the other hand, much more difficult
many--body problem. In the following we always assume that the quasiparticle
energies are necessarily approximated by the LDA energy bands.

In the framework of the density-functional linear-response method the
problem of calculating the phonon spectra and the electron--phonon
interaction is reduced to finding the first-order variations in the
one-electron wave functions, the charge density and the effective potential
induced by the presence of a phonon with a given wave vector ${\bf q.}$
These quantities are connected by the so-called Sternheimer equation, which
is the Schr\"odinger equation to linear order (we use atomic Rydberg units
throughout the paper):

\begin{equation}
(-\nabla ^2+V_{eff}-\epsilon _{{\bf k}j})\frac{\delta ^{\pm }\psi _{{\bf k}j}%
}{\delta R_\mu }+(\frac{\delta ^{\pm }V_{eff}}{\delta R_\mu }-\frac{\delta
^{\pm }\epsilon _{{\bf k}j}}{\delta R_\mu })\psi _{{\bf k}j}=0,  \label{E1}
\end{equation}
where $\epsilon _{{\bf k}j},\psi _{{\bf k}j}$ are the unperturbed energies
and wave functions of vector ${\bf k}$ and band $j$; $V_{eff}$ is the
effective potential of the DFT; $\delta ^{\pm }\epsilon _{{\bf k}j}/\delta
R_\mu ,\delta ^{\pm }\psi _{{\bf k}j}/\delta R_\mu ,\delta ^{\pm
}V_{eff}/\delta R_\mu $ are the first-order changes in these quantities
induced by the displacements of atoms in the positions ${\bf R}+{\bf t}$ ($%
{\bf R}$ is the basis vector and ${\bf t}$ is the translation vector) moving
in the direction $\mu $ according to: $\delta {\bf t}_R=\delta {\bf A}%
_Rexp(+i{\bf qt})+\delta {\bf A}_R^{*}exp(-i{\bf qt})$ where $\delta {\bf A}%
_R$ is the complex polarization vector. The sign ''$\pm $'' in the operator $%
\delta ^{\pm }/\delta R_\mu $ refers to the changes associated with the
contributions to $\delta {\bf t}_R$ which represent two travelling waves of
vectors $+{\bf q}$ and $-{\bf q.}$ Note that $\delta ^{\pm }\epsilon _{{\bf k%
}j}/\delta R_\mu =\langle {\bf k}j|\delta ^{\pm }V_{eff}/\delta R_\mu |{\bf k%
}j\rangle $ vanishes unless ${\bf q}=0.$

Equation (\ref{E1}) must be solved self-consistently since the induced
charge density $\delta ^{\pm }\rho /\delta R_\mu $ expressed via $\delta
^{\pm }\psi _{{\bf k}j}/\delta R_\mu $ screens out the external perturbation 
$\delta ^{\pm }V_{ext}/\delta R_\mu .\,$The latter is simply given by the
change in the bare Coulomb potential of the nuclei. We thus see that the
scheme is the linearized analog of the original Kohn-Sham equations. If Eq.(%
\ref{E1}) is solved and the value of $\delta ^{\pm }\rho /\delta R_\mu $ is
available, the phonon frequencies $\omega _{{\bf q}\nu }$ and the
eigenvectors ${\bf \eta }_{{\bf q}\nu }$ ($\nu $ numerates phonon branches)
are obtained by diagonalizing the adiabatic dynamical matrix of the crystal.
The latter is the second-order derivative of the DFT total energy with
respect to $\delta R_{\mu \text{ }}$and $\delta R_{\mu ^{\prime }}^{\prime }$
and is conveniently represented by the Hellmann-Feynman theorem. In
practice, however, there is a problem in using the Hellmann-Feynman
expression connected with the fact that the unperturbed wave functions are
obtained not as exact solutions of the Kohn--Sham equations but by expanding
them into some basis set $\chi _\alpha ^{{\bf k}}$:

\begin{equation}
\psi _{{\bf k}j}=\sum_\alpha \chi _\alpha ^{{\bf k}}A_\alpha ^{{\bf k}j}
\label{E2}
\end{equation}
Applying the Rayleigh-Ritz variational principle, this leads to finding the
unknown coefficients $A_\alpha ^{{\bf k}j}$ from the matrix eigenvalue
problem: 
\begin{equation}
\sum_\alpha \langle \chi _\beta ^{{\bf k}}|-\nabla ^2+V_{eff}-\epsilon _{%
{\bf k}j}|\chi _\alpha ^{{\bf k}}\rangle A_\alpha ^{{\bf k}j}=0  \label{A1}
\end{equation}
The consequence of the variational formulation is that, in general, the
Hellmann--Feynman contribution is supplemented with the correction expressed
via the derivatives of the basis functions with respect to the
displacements, $\delta ^{\pm }\chi _\alpha ^{{\bf k}}/\delta R_\mu $, which
accounts for the incompleteness of the basis. The correction enters as the
matrix element: 
\begin{equation}
\langle \sum_\beta A_\beta ^{{\bf k}j}\frac{\delta ^{\pm }\chi _\beta ^{{\bf %
k}}}{\delta R_\mu }|-\nabla ^2+V_{eff}-\epsilon _{{\bf k}j}|{\bf k}j\rangle 
\label{A2}
\end{equation}
and is obviously not equal to zero if the states $\psi _{{\bf k}j}$ are only
variationally accurate. The IBS corrections explicitly disappear for the
plane-wave basis set but must be taken into account for the basis of linear
muffin-tin orbitals which is used in our work.

To treat properly the changes in the wave functions caused by variation of
some external parameters one should check that the basis functions are well
adjusted in the whole region of the parameters variations. Within the linear
response, this is achieved by representing the first-order perturbations in $%
\psi _{{\bf k}j}$ in the form: 
\begin{equation}
\frac{\delta ^{\pm }\psi _{{\bf k}j}}{\delta R_\mu }=\sum_\alpha (\chi
_\alpha ^{{\bf k}}\frac{\delta ^{\pm }A_\alpha ^{{\bf k}j}}{\delta R_\mu }+%
\frac{\delta ^{\pm }\chi _\alpha ^{{\bf k}}}{\delta R_\mu }A_\alpha ^{{\bf k}%
j})  \label{E3}
\end{equation}
where $\delta ^{\pm }A_\alpha ^{{\bf k}j}/\delta R_\mu $ are the changes in
the expansion coefficients. This is advantageous comparing to the standard
perturbation theory, where only the unperturbed states are used as a basis
for representing $\delta ^{\pm }\psi _{{\bf k}j}/\delta R_\mu $ [This is
described by only the first contribution in (\ref{E3})]. The whole expansion
(\ref{E3}) must be fastly convergent because the change $\delta ^{\pm }\chi
_\alpha ^{{\bf k}}/\delta R_\mu $ in the LMTO basis is tailored to the
perturbation just like the original LMTO basis $\chi _\alpha ^{{\bf k}}$ is
tailored to the one--electron potential.

In order to find the coefficients $\delta ^{\pm }A_\alpha ^{{\bf k}j}/\delta
R_\mu $ we first construct the functional of the dynamical matrix,
minimization of which with respect to $\delta ^{\pm }\psi _{{\bf k}j}/\delta
R_\mu $ leads to the equation (\ref{E1}), and then apply the Rayleigh-Ritz
variational principle with the trial functions represented by Eq.\ref{E3}.
This leads to a set of matrix equations for $\delta ^{\pm }A_\alpha ^{{\bf k}%
j}/\delta R_\mu .\,$ The detailed description of this method as well as its
applications for calculating phonon dispersions can be found in Refs. %
\onlinecite{LR,UFN,LRPRB}.

In this paper we test the produced self-consistent change in the
one-electron potential as the potential of electrons interacting with the
phonon mode $\omega _{{\bf q}\nu }$. Our task is to evaluate the
electron-phonon matrix element $g_{{\bf k}+{\bf q}j^{\prime },{\bf k}j}^{%
{\bf q}\nu }$. The latter is conventionally written in the form: 
\begin{equation}
g_{{\bf k}+{\bf q}j^{\prime },{\bf k}j}^{{\bf q}\nu }=\langle {\bf k}+{\bf q}%
j^{\prime }|\delta ^{{\bf q}\nu }V_{eff}|{\bf k}j\rangle   \label{E4}
\end{equation}
where both states $\psi _{{\bf k}j\text{ }}$and $\psi _{{\bf k}+{\bf q}%
j^{\prime }}$ have the Fermi energy $\epsilon _F$ and where the change in
the potential is transformed from the Cartesian system to the system
associated with the eigenvectors $\eta _{{\bf q}\nu }(R\mu )$ of a
particular ${\bf q}\nu $ -- mode:

\begin{equation}
\delta ^{{\bf q}\nu }V_{eff}=\sum_{R,\mu }\frac{\eta _{{\bf q}\nu }(R\mu )}{%
(M_R\omega _{{\bf q}\nu })^{1/2}}\times \frac{\delta ^{+}V_{eff}}{\delta
R_\mu }  \label{E5}
\end{equation}
($M_{R\text{ }}$are the nuclei masses).

It is not obvious but the expression (\ref{E4}) for $g_{{\bf k}+{\bf q}%
j^{\prime },{\bf k}j}^{{\bf q}\nu }$ should be corrected for the
incompleteness of the basis functions. This fact immediately follows if we
will interpret the matrix element (\ref{E4}) as a splitting of the
degenerate band $\epsilon _{{\bf k}j}=\epsilon _{{\bf k}+{\bf q}j^{\prime
}}=\epsilon _F$ due to the phonon distortion. $\,$ Consider first a
non-degenerate state for which the first--order correction to the eigenvalue 
$\epsilon _{{\bf k}j}$ is found as a change in the eigenvalue of the matrix
problem (\ref{A1}). It is given by

\begin{eqnarray}
&&\delta \epsilon _{{\bf k}j}=\langle {\bf k}j|\delta V_{eff}|{\bf k}%
j\rangle +  \nonumber \\
&&\langle \sum_\beta A_\beta ^{{\bf k}j}\delta \chi _\beta ^{{\bf k}%
}|-\nabla ^2+V_{eff}-\epsilon _{{\bf k}j}|{\bf k}j\rangle +  \nonumber \\
&&\langle {\bf k}j|-\nabla ^2+V_{eff}-\epsilon _{{\bf k}j}|\sum_\alpha
A_\alpha ^{{\bf k}j}\delta \chi _\alpha ^{{\bf k}}\rangle  \label{E6}
\end{eqnarray}
where $\delta \chi _\alpha ^{{\bf k}}$ and $\delta V_{eff}$ denote the
changes in the basis functions and the potential due to some change in the
external parameters of the hamiltonian. This formula contains both the
expression of the standard perturbation theory (first term here) as well as
incomplete-basis-set corrections [second and third contributions in (\ref{E6}%
)]. It would be advantageous to use this formula because the eigenvalues of
the matrix problem (\ref{A1}) are variationally accurate for the whole range
of parameters variation. The same valid if we develop a perturbation theory
for the change in the degenerate band $\epsilon _{{\bf k}j}=\epsilon _{{\bf k%
}+{\bf q}j^{\prime }}\,$ within (\ref{A1}). This leads to the result:

\begin{eqnarray}
&&g_{{\bf k}+{\bf q}j^{\prime },{\bf k}j}^{{\bf q}\nu }=\langle {\bf k}+{\bf %
q}j^{\prime }|\delta ^{{\bf q}\nu }V_{eff}|{\bf k}j\rangle +  \nonumber \\
&&\langle \sum_\alpha \delta ^{{\bf q}\nu }\chi _\alpha ^{{\bf k}-{\bf q}%
}A_\alpha ^{{\bf k}+{\bf q}j^{\prime }}|-\nabla ^2+V_{eff}-\epsilon _{{\bf k}%
j}|{\bf k}j\rangle +  \nonumber \\
&&\ \langle {\bf k}+{\bf q}j^{\prime }|-\nabla ^2+V_{eff}-\epsilon _{{\bf k}+%
{\bf q}j^{\prime }}|\sum_\alpha \delta ^{{\bf q}\nu }\chi _\alpha ^{{\bf k}%
}A_\alpha ^{{\bf k}j}\rangle  \label{E7}
\end{eqnarray}
where $\delta ^{{\bf q}\nu }\chi _\alpha ^{{\bf k}}\,\,$denotes the
variation of the MT basis functions due to the phonon distortion of the $%
{\bf q}\nu $ --mode and is connected with the variation $\delta ^{{\bf +}%
}\chi _\alpha ^{{\bf k}}/\delta R_\mu $ in the same way as for the induced
potential, Eq. (\ref{E5}). The last two contributions in (\ref{E7})
represent the IBS--corrections which are not vanished unless $\psi _{{\bf k}%
j},\psi _{{\bf k}+{\bf q}j^{\prime }}$ are the exact solutions.

Formally, the expression (\ref{E7}) can also be obtained by repeating the
standard quantum-mechanical derivation of the Fermi ''golden rule'' for the
wave functions represented in terms of the basis set according to (\ref{E2}%
). The matrix element of the electron-phonon interaction is obtained by
considering the scattering rate for transitions from an initial, unperturbed
state $\psi _{{\bf k}j}(t)$ into a final perturbed state $\tilde \psi _{{\bf %
k}^{\prime }j^{\prime }}(t)$ at the time moment $t\,$which is given by the
overlap integral squared:

\begin{equation}
|\langle {\bf k}j(t)|\widetilde{{\bf k}^{\prime }j^{\prime }(t)}\rangle |^2
\label{E8}
\end{equation}
Since the final state corresponds to the displaced lattice, the best
variational estimate for $\tilde \psi _{{\bf k}^{\prime }j^{\prime }}(t)$
must include the orbitals centered at the new atomic positions and adjusted
to the new one-electron potential. To linear--order with respect to the
displacements this leads to finding the change in the basis $\chi _\alpha ^{%
{\bf k}}\,$ which gives rise to the IBS-corrections entered (\ref{E7}).

The expression (\ref{E7}) is thus the linear--response analogy of evaluating
the electron--phonon matrix elements via the splitting of the bands in the
frozen--phonon supercell method as done in Ref. \onlinecite{Liecht}. It is
less sensitive to the errors in the wave functions introduced by the
variational principle, has a correct long--wavelength behavior and allows
one to avoid the inclusion of $d-f$ transitions in $d-$electron systems.

For the electron-phonon spectral distribution functions $\alpha ^2F(\omega )$
we employ the expression\cite{Allen} in terms of the phonon linewidths $%
\gamma _{{\bf q}\nu }$

\begin{equation}
\alpha ^2F(\omega )=\frac 1{2\pi N(\epsilon _F)}\sum_{{\bf q}\nu }\frac{%
\gamma _{{\bf q}\nu }}{\omega _{{\bf q}\nu }}\delta (\omega -\omega _{{\bf q}%
\nu }),  \label{E9}
\end{equation}
where $N(\epsilon _F)$ is the electronic density of states per atom and per
spin at the Fermi level. When the energy bands around the Fermi level are
linear in the range of phonon energies, the linewidth is given by the Fermi
''golden rule'' and is written as follows

\begin{equation}
\gamma _{{\bf q}\nu }=2\pi \omega _{{\bf q}\nu }\sum_{{\bf k}jj^{\prime
}}|g_{{\bf k+q}j^{\prime },{\bf k}j}^{{\bf q}\nu }|^2\delta (\epsilon _{{\bf %
k}j}-\epsilon _F)\delta (\epsilon _{{\bf k}+{\bf q}j^{\prime }}-\epsilon _F).
\label{E10}
\end{equation}
The spectral distribution function (\ref{E9}) and its first reciprocal
moment $\lambda $ are usually used to describe such important manifestation
of the EPI as superconductivity and some normal-state properties. One of
such properties is an enhancement of the electronic mass for the electron at
the Fermi energy when its velocity $v_{{\bf k}}$ is reduced by the factor $%
1+\lambda _{{\bf k}}$ due to the interaction with phonons. The value of $%
\lambda _{{\bf k}}$ is given by the reciprocal moment of the so-called ${\bf %
k-}$dependent electron-phonon spectral function $\alpha _{{\bf k}}^2F(\omega
).\,$This renormalization is observed in the de Haas-van Alphen and
cyclotron--resonance experiments. As its consequence, the low-temperature
electronic specific heat is also renormalized. For the latter effect it is
sufficient to know only the Fermi--surface averaged value of $\lambda _{{\bf %
k}}$, i.e. $\lambda $.

Full description of the superconducting state can be obtained by solving the
Eliashberg gap equations which relate the energy--gap function and the
renormalization parameter for superconducting state to the electron-phonon
and the electron-electron interactions in the normal state\cite{Eliashberg}.
The electron-phonon-coupling function is given by $\alpha ^2F(\omega )$. The
Coulomb interaction is usually represented by some constant $\mu ^{*}$.
Detailed nature of the effective Coulomb repulsion is not very well known.
Fortunately, various definitions for $\mu ^{*}$ have only a weak influence
on the solution of the gap equations and its values can, {\it e.g.}, be
found by adjusting the calculated transition temperatures to the
experimental ones.

Electron-phonon scattering has a dominant contribution to the electrical
resistivity for reasonably pure metals except very low--temperature region
where the impurity and electron--electron scattering are important. The
influence of the EPI on the transport properties are described in terms of
the transport spectral function\cite{AllenTR} $\alpha _{tr}^2F(\omega
)=\alpha _{out}^2F(\omega )-\alpha _{in}^2F(\omega )$ where:

\begin{eqnarray}
&&\alpha _{out(in)}^2F(\omega )=  \nonumber \\
&&\frac 1{N(\epsilon _F)\langle v_x^2\rangle }\sum_\nu \sum_{{\bf k}j{\bf k}%
^{\prime }j^{\prime }}|g_{{\bf k}^{\prime }j^{\prime },{\bf k}j}^{{\bf k}%
^{\prime }-{\bf k}\nu }|^2v_x({\bf k})v_x({\bf k}^{(^{\prime })})\times  
\nonumber \\
&&\ \ \delta (\epsilon _{{\bf k}j}-\epsilon _F)\delta (\epsilon _{{\bf k}%
^{\prime }j^{\prime }}-\epsilon _F)\delta (\omega -\omega _{{\bf k}^{\prime
}-{\bf k}\nu })  \label{E12}
\end{eqnarray}
Here $\langle v_x^2\rangle $ is the average square of the $x$ component of
the Fermi velocity. In the lowest--order variational approximation (LOVA)\
for the solution of the Boltzman equation the expressions for electrical and
thermal resistivities are

\begin{equation}
\rho (T)=\frac{\pi \Omega _{cell}k_BT}{N(\epsilon _F)\langle v_x^2\rangle }%
\int\limits_0^\infty \frac{d\omega }\omega \frac{x^2}{sinh^2x}\alpha
_{tr}^2F(\omega ),  \label{E13}
\end{equation}
\begin{eqnarray}
&&w(T) =\frac{6\Omega _{cell}}{\pi k_BN(\epsilon _F)\langle v_x^2\rangle }%
\int\limits_0^\infty \frac{d\omega }\omega \frac{x^2}{sinh^2x}\times 
\nonumber \\
&&\lbrack \alpha _{tr}^2F(\omega )+\frac{4x^2}{\pi ^2}\alpha
_{out}^2F(\omega )+\frac{2x^2}{\pi ^2}\alpha _{in}^2F(\omega )]  \label{E14}
\end{eqnarray}
with $x=\omega /2k_BT$. The LOVA results (\ref{E12}), (\ref{E13}) give the
upper bound to the resistivities and allow us to test the calculated $\alpha
_{tr}^2F(\omega )$.

\smallskip\ 

{\bf III. RESULTS.}

\smallskip\ 

{\bf a. Technicalities.}

\smallskip\ 

Our calculations of phonon dispersions and electron-phonon interactions for
the elemental metals such as fcc-- $Al,Cu,Pb,Pd,$ and bcc-- $Mo,Nb,Ta,V$ are
performed in the framework of the linear--response LMTO method\cite{LRPRB}.
The details of the calculations are the following: We find the dynamical
matrix and the phonon linewidths for these materials as a function of wave
vector for a set of irreducible ${\bf q}$--points at the $(8,8,8)$%
--reciprocal lattice grid [29 points per $1/48$th part of the Brillouin zone
(BZ)]. The $(I,J,K)$ reciprocal lattice (or Monkhorst--Pack\cite{Monkhorst})
grid is defined in a usual manner: ${\bf q}_{ijk}=\frac iI{\bf G}_1+\frac jJ%
{\bf G}_2+\frac kK{\bf G}_3,$ where ${\bf G}_1,{\bf G}_2,{\bf G}_3$ are the
primitive translations in the reciprocal space. The self--consistent
calculations performed for every wave vector involve the following
parameters: We use $3\kappa -spd-$LMTO\ basis set (27 orbitals) with the
one--center expansions performed inside the MT--spheres up to $l_{max}=6.$
In the interstitial region the basis functions are expanded in plane waves
up to the cutoff approximately corresponding to 70, 140, and 200 plane waves
per $s,p,$ and $d-$orbitals, respectively. All semicore states lying higher
than $-4Ry$ are treated as valence states in separate energy windows. The
induced charge densities and the screened potentials are represented inside
the MT--spheres by spherical harmonics up to $l_{max}=6$ and in the
interstitial region by plane waves with the cutoff corresponding to the $%
(16,16,16)$ fast--Fourier--transform grid in the unit--cell of direct space.

The ${\bf k}$--space integration needed for constructing the induced charge
density and the dynamical matrix is performed over the $(16,16,16)$ grid
(145 points per $1/48$th part of the BZ) which is twice denser than the grid
of the phonon wave vectors ${\bf q}.$ We use the improved tetrahedron method
of Ref. \onlinecite{TETR2}. However, the integration weights for the ${\bf k-%
}$points at this $(16,16,16)$ grid have been found to take precisely into
account the effects arising from the Fermi surface and the energy bands.
This is done with help of the energies $\epsilon _{{\bf k}j}$ generated by
the original full--potential LMTO\ method at the $(32,32,32)$ grid (897
points per $1/48$ BZ). The procedure is in details explained in Ref. %
\onlinecite{LRPRB} and allows us to obtain more convergent results with
respect to the number of ${\bf k}$ points.

The ${\bf k}$--space integration for the phonon linewidths $\gamma _{{\bf q}%
\nu }$ is very slowly convergent because it involves two $\delta $ functions
according to Eq. (\ref{E10}). It is performed with help of the $(32,32,32)$
grid in the BZ by means of the tetrahedron method of Ref. \onlinecite{TETR1}%
. The largest numerical error of $\alpha ^2F(\omega )$ comes from the
integration over ${\bf q}$ in the expression (\ref{E9}). Its magnitude, we
estimated by performing the integration over merely the band--structure
factor [which is $\gamma _{{\bf q}\nu }$ approximated by $\sum_{{\bf k}%
jj^{\prime }}\delta (\epsilon _{{\bf k}j}-\epsilon _F)\delta (\epsilon _{%
{\bf k}+{\bf q}j^{\prime }}-\epsilon _F)$] using respectively $(8,8,8)$ and $%
(32,32,32)$ grids and found to be not larger than $7\%$ in all cases.

A few words should be said about the lattice parameters used in the
calculations. It is known that the equilibrium cell volume $V$ found
theoretically by the corresponding LDA--based total--energy calculation is
frequently obtained slightly lower than the experimental volume $V_0$. This
usually leads to the calculated at $V_0$ phonon frequencies which are softer
comparing to the experimental ones. Often, a better agreement with the
experiment can be obtained by performing the linear--response calculations
at the theoretical volume. This is, in principle, a justified procedure from
theoretical point of view. Unfortunately, the prescription does not work
when calculating the phonon linewidths and $\alpha _{tr}^2F(\omega ):$ the
results of calculated electrical and thermal resistivities agree less well
with the experiment. The reason for this discrepancy is connected with the
sensitivity of these quantities to the shape of the Fermi surface. It turns
out that the use of the Fermi surfaces calculated at the experimental
lattice constants considerably improves the results. We can thus use
theoretical volumes in the linear--response calculations of phonon
dispersions and the electron--phonon matrix elements. To find the phonon
linewidths and $\alpha _{tr}^2F(\omega )$ we can use, on the other hand, the
energy bands entered (\ref{E10}) which are generated at the experimental
lattice constants. We understand that it is not well justified procedure to
use different lattice parameters in one calculation, but it somewhat helps
to minimize the errors connected with the LDA by simple means. The actual
volume ratios $V/V_0$ used in our calculations to find the changes in the
one--electron potentials are listed in Table I.

Another comment concerns the choice of the exchange--correlation potential.
The general strategy employed by us is to use the exchange--correlation
formula which gives the best prediction of the cell volumes. The
Barth--Hedin--like formula after Ref. \onlinecite{Moruzzi} is employed for
all the metals except $Cu$ and $V.$ For the $3d$--metals we have found that
this formula gives the theoretical volumes which are too small ($V/V_0\sim
0.9$). For $Cu$ and $V$ the Ceperley--Alder form\cite{Ceperley} of the
exchange--correlation potential parametrized after Ref. \onlinecite{Vosko}
is used which gives the ratios $V/V_0$ being much closer to unity (see Table
I).

Finally note that in the previous publications\cite{LR,UFN,EPI} we have used
a different method for treating the full--potential terms in the calculation
which was based on the atomic cells and the one--center spherical--harmonic
expansions\cite{Forces}. This method is not directly applicable to calculate
phonon dispersions for materials with open structures such, e.g., the
diamond structure and requires the replacement of empty sites of the lattice
by empty spheres. This complicates the evaluation of the dynamical matrix.
In recent publication\cite{LRPRB} we have employed another approach based on
the plane--wave expansions for the LMTOs in the interstitial region and have
applied the method to calculate the phonon spectra in $Si$ and $NbC.$ While
the materials considered in this work have close--packed bcc or fcc
structures we also apply this new method which is more general for practical
use. Some of the results for $Al$, $Nb,$ and $Mo$ previously published\cite
{LR,UFN,EPI}  do not noticeably differ from those presented below in this
paper.

\smallskip\ 

{\bf b. Lattice-dynamical properties.}

\smallskip\ 

In Table I we report the values of the calculated phonon frequencies at the
high--symmetry points $X$ and $L$ for the fcc metals $Al,Cu,Pb$ and $Pd$ as
well as at the points $H$ and $N$ for the bcc metals $Mo,Nb,Ta,$ and $V$ .
For the comparison, the experimental frequencies\cite{Landolt} are also
listed in Table I along with the theoretical--to--experimental volume ratios
which have been used in the calculations.

Our results for the phonon dispersions along several symmetry directions
together with the corresponding densities of states for these materials are
displayed in Fig.1(a)-(h). The theoretical lines result from the
interpolation between the calculated frequencies which are denoted by
circles. Many neutron--diffraction measurements are available\cite{Landolt}
for nearly all the metals considered here and these data are also shown in
Fig.1 by triangles. The only exception is $V$ for which the dispersion
relations cannot be studied with neutrons since $V$ is an almost totally
incoherent neutron scatterer. While some $X$--ray diffraction measurements
exist in the literature\cite{Landolt} their accuracy seems to be less
satisfactory than corresponding neutron--scattering data for other materials
and we do not show the experimental points for $V.$

From Fig.1 we see that the agreement between theory and experiment is good.
Most of the calculated frequencies agree within a few percent with those
measured. This also follows from the numerical values listed in Table I. In
particular, for $Al$ [Fig.1(a)] a very good agreement is found in all the
directions. As we have mentioned already, this calculation is performed at
the theoretical volume $(V/V_0=0.955)$. We have also checked the set-up with
the experimental volume and found a considerable softening (about $20\%$) of
the transverse modes. This illustrates the importance of performing
lattice-dynamical calculations at the theoretical volumes.

The most important consequence of our calculation for lead [Fig.1(b)] is
that the pronounced dip of both the longitudinal and transverse branches
near the $X$--point is well reproduced . We have also found a slight
overestimate of the transverse phonon frequencies near this zone boundary
which can be attributed neither to the discrepancy in the cell volume ($%
V/V_0=1.002$) nor to the neglection of the semicore states. (We have in fact
included both $5d$, and $6d$--states in the main valence panel). This kind
of disagreement has also been recently reported in Ref. %
\onlinecite{Gironcoli} using the linear--response pseudopotential technique.
It is possibly connected with the use of the local density approximation or
the lack of spin--orbit coupling effects in our calculation.

The most interesting cases are $V,Nb,$ and $Ta$ [Fig.1(c),(d),(e),
respectively]. The materials belong to Group V of the periodic table and all
of them show anomalious behavior of the phonon dispersion curves. The
presence of anomalies is, first of all, connected with the well--known dip
of the longitudinal mode in the $(00\xi )$ direction. The dip is correctly
reproduced by our calculation. Another important features of our calculation
are (i) the softening of the transverse mode along the $(00\xi )$ direction
at long wavelengths which is rather sharp in both $V,Nb$ and is weaker in $Ta
$ as well as (ii) the crossover of two transverse branches in the $(\xi \xi
0)$ direction. The latter is in fact predicted for $Ta\,$ because the
measured dispersions in this direction are absent for the $T_1$--branch
except the zone--boundary point $N$.

The theoretical phonon dispersions for $Mo$ also agree well with the
experiment. The consequences of our calculations here are the reproduced
softening near the $H$ point and the absence of the large dip along the $%
(\xi \xi \xi )$ direction near $\xi \sim 0.7.$ The dip is presented in the
dispersion curves of nearly all bcc metals except of $Mo$ and $Cr$, and is
certainly the feature of the behavior of the crystalline structure factor
which enters the dynamical matrix as a sum over lattice vectors with the
phase shift $exp(i{\bf qt}).$ Its absence indicates a considerable
wave--vector dependence of the electron--phonon matrix elements which seems
to be well reproduced by our method.

Finally we compare the results of our calculations for $Cu$ and $Pd\,$ which
are presented in Fig.1(f) and Fig.1(g). The dispersion relations for these
materials do not show the anomalies and are smooth. A slight overestimation
of the theoretical phonon frequencies is found for both these metals . The
overestimation can likely be corrected by the use of the cell volumes in the
calculation which would be slightly larger than those found within the LDA.
However, such a procedure is not justified theoretically. We think that
employing the gradient--corrected density functional\cite{GGA} known to
predict much better the equilibrium lattice parameter will allow us to
improve these results.

\smallskip\ 

{\bf c. Superconducting properties.}\ 

\smallskip\ 

We now discuss applications of our {\it ab initio} linear--response method
to calculate the electron--phonon coupling and superconducting--state
properties. First, we present our results for the spectral distribution
functions and $\lambda $. Second, we describe our applications to solving
the Eliashberg gap equations.

Calculated $\alpha ^2F(\omega )$ for $Al$ is shown in Fig.1(a) (full line).
The positions of the maxima here are conditioned by the form of the phonon
density of states with the low-frequency phonon peak suppressed by the
coupling function $\alpha ^2(\omega )$ (dashed line). The broad phonon
spectrum in $Al$ is extended up to the maximal frequency $\omega
_{max}\approx 470K.$ The theoretical $\alpha ^2F(\omega )$ in Fig.1(a) is
compared with the results of the tunneling measurements\cite{Wolf}
(squares). We find a rather good agreement between the two curves. In fact,
our $\alpha ^2F(\omega )$ is also found to be practically identical to the
empirical pseudopotential result of Ref. \onlinecite{Carbotte} based on the
rigid--ion approximation. The latter is known to work well in simple metals.
General agreement is found between our and the {\it ab initio}
frozen--phonon results of Dacorogna {\it et}. al.\cite{Dacorogna} for the
dispersion of the phonon linewidths along the high--symmetry directions. The
only exception is that, in the $(00\xi )\,$ direction, our longitudinal
branch of $\gamma _{{\bf q}\nu }$ exceeds theirs by a factor of $2$. This is
presumably connected with replacing the $\delta $ functions in (\ref{E10})
by Gaussians used in Ref. \onlinecite{Dacorogna}. However, the relative
weight of our high $\gamma $ values in the integrated quantities, such as $%
\alpha ^2F(\omega )$ and $\lambda ,$ is found to be very small. Our value of 
$\lambda $ is $0.44$ which is very close to the value of $\lambda
_{tun}=0.42 $ extracted from the tunnelling measurements\cite{Wolf}. The
frozen--phonon \cite{Dacorogna} and linear--response calculations of Winter%
\cite{Winter} gave, respectively, $\lambda =0.45\,$ and $0.38.$ $\,$The
value of $\lambda _{s-h}$ extracted from the electronic specific-heat
coefficient $\gamma $ and our calculated density of states $N(\epsilon _F)$
using the relation

\begin{equation}
1+\lambda _{s-h}=\frac{3\gamma }{2\pi ^2k_BN(\epsilon _F)},  \label{S3}
\end{equation}
is $0.43$ (see Table II). In order to check previous conclusions\cite
{Winter,Papa} about the inapplicability of the RMTA for $sp$ metals, we also
performed such a calculation and indeed found $\lambda _{RMTA}=0.14.$

Lead is a well--studied classical example of strong--coupled superconductor
with $T_c=7.19K$ and its tunnelling spectra have been studied a long time ago%
\cite{Lead}. Obtained $\alpha ^2F(\omega )$ using our linear--response
method is presented in Fig.2(b) where it is compared with the results of the
measurements\cite{Lead}. The two curves are similar. Our calculated $\lambda
=1.68$ is found to be $8\%$ larger than the tunnelling value $1.55$ and only 
$2\%$ larger than the value $1.64$ extracted from specific--heat data (see
Table II). This disagreement is well within the accuracy of our calculation.

We now report our results for $V,Nb$ and $Ta$ which are the best--studied
elemental superconductors because of their relatively high--$T_c\,$values.
Especially, for $Nb$ which has the highest $T_c=9.25K$ among the elemental
metals, there exist many experimental investigations of the tunneling spectra%
\cite{Bostock,Robinson,Arnold,Bostock1} and theoretical RMTA--based
calculations\cite{Papa,Butler,Harmon}. Unfortunately, some of the results
which have been reported in the literature are controversial. First, the
RMTA calculations give the values of $\lambda \,$ varying from $1.12$ to $%
1.86$. Second, the tunneling estimates of the coupling constant for $Nb$ and 
$V$ are difficult because of the oxidation of the surface layers. Having
lower transition temperature, such oxides act on the tunneling spectra due
to the proximity effect. The experimental $\lambda $ in $Nb$ varied in the
past from the values\cite{Bostock} $0.58-0.68$ with negative or anomaliously
small $\mu ^{*}$ to the value\cite{Robinson} $0.9$. At present, a
satisfactory explanation of the anomalious behavior of the thermally
oxidized tunnelling junctions in $Nb$ appears to be possible which gives the
values\cite{Arnold,Bostock1} of $\lambda _{tun}=0.92\div 1.22.$

In Fig.2(c),(d),(e) we present our calculated $\alpha ^2F(\omega )$ (full
lines) for $V,Nb$ and $Ta,\,$respectively. They all have rather broad
spectra extended up to $\omega _{max}\approx 370K,310K,\,$ and $240K$. The
coupling function $\alpha ^2(\omega )$ (dashed line) in these metals only
slightly deviates from constant in major part of the frequencies. The
approximation $\alpha ^2=const.$ works very well in $Nb$ and $Ta$. This
qualitative result implies that the electron-phonon coupling can be
factorized into electronic and phonon--dependent factors\cite{McMillan,RMTA}.

In Fig.2 the calculated spectral functions are compared with the results of
the tunneling measurements (squares). As it can be seen for $V$ [Fig.2(c)],
our $\alpha ^2F(\omega )$ disagrees with the measured one\cite{Zasadzinski}
because of the appearance of the upper phonon peak not presented in the
experiment. Even though the theoretical $\alpha ^2F(\omega )$ should be
broadened because the $\delta $ function in Eq. (\ref{E9}) ought to be a
Lorentzian of half--width $\gamma _{{\bf q}\nu }$, the electron--phonon
coupling estimated by us ($\lambda =1.19$) is $40-50\%$ stronger than the
obtained $\lambda _{tun}=0.82$. The same situation is found for $Nb.$ Our
calculation here [Fig.2(d)] also does not show the suppression of the
longitudinal peak. The latter is absent in nearly all the experiments for
this metal\cite{Arnold}$.$ [A typical measured spectrum\cite{Wolf} is shown
in Fig.2(d) by squares]. As a result, the calculated $\lambda =1.26$ is $20\%
$ higher than $\lambda _{tun}=1.04$\cite{Wolf}$.$ The discrepancy found by
us has already been reported in the past RMTA--based calculations\cite
{Butler,Harmon}. To check the consistency of our results with the earlier
ones, we have performed our own RMTA calculations and obtained complete
agreement between them. We thus conclude that the full inclusion of
screening does not resolve the problem of the suppressed longitudinal peak.

Unfortunately, our comparison with the experiment is complicated by the
proximity effect and the extraction of the tunneling densities of states
depends on the way how the measured data are processed. For example, in $Nb$
the value of $\lambda _{tun}=1.22$ deduced from the tunneling experiments
(which is only $3\%$ lower than that found by us) has been reported in the
literature\cite{Bostock1}. The obtained $\alpha ^2F(\omega )$ [denoted in
Fig.2(d) by triangles] is found much closer to our calculation.

A better understanding of the present situation can be achieved by comparing
the theoretical and the experimental\cite{Shen,Wolf} tunneling spectra for $%
Ta$ since its superconducting properties are close to those of $V$ and $Nb$
but this metal is much less reactive with oxygen. Such a comparison is given
in Fig.2(e). We find rather good agreement between the both curves. In
particular, the upper phonon peak is not suppressed in the measured $\alpha
^2F(\omega )\,$and its amplitude is comparable with that calculated by us.
As a result, the theoretical $\lambda =0.86$ agrees within $10\%$ with the $%
\lambda _{tun}=0.78.$

From view of the data on $Ta$ we cannot consider the discrepancy found for $V
$ and $Nb$ as a drawback of either our linear--response method or the use of
the local density approximation. Partially this conclusion is also verified
by alternative estimates of the coupling constant based on the
specific--heat data and the de Haas--van Alphen (dHvA) experiments, but it
should be noted that both cyclotron masses and the specific--heat
coefficient are also enhanced by the electron--electron interactions.
Evaluation of the average coupling from the specific--heat measurements\cite
{SH} yields an enhancement of $1.00$ for $V$ and $1.17$ for $Nb$ (see Table
II). If one uses a more recent value\cite{Vedeneev} of $\gamma $ for $V$
rather than listed in Ref. \onlinecite{SH} one obtains the enhancement $%
\lambda _{s-h}=$ $1.17$ which is nearly coincides in this metal with our $%
\lambda =1.19$ . Comparison between the LDA band masses with those measured
by dHvA effect\cite{dHvA} yields the enhancement of $1.33$ for $Nb$, which
is close to the value $1.26$ found in our calculation. Another important
result is that the measured {\em variation} of the mass enhancement for the
various cyclotron orbits in $Nb$ agrees also well with our calculation.
Namely, we have found a decomposition of $\lambda $ by the Fermi--surface
sheets: (i) octahedron, (ii) jungle gym and (iii) ellipsoids, and have
obtained the contributions $\lambda ^{\text{i}}=1.44,\lambda ^{\text{ii}%
}=1.37,$ and $\lambda ^{\text{iii}}=1.08.$ They can be compared with the
measured $\lambda _{dHvA}^{\text{i}}=1.71,\lambda _{dHvA}^{\text{ii}}=1.43,$
and $\lambda _{dHvA}^{\text{iii}}=1.10.$ (Note that the calculated within
the RMTA anisotropy of the mass enhancement strongly disagrees with these
data\cite{dHvA}.) Moreover, we have estimated the transport constants $%
\lambda _{tr}$ for these metals both from the calculated and measured
resistivity data. The values of $\lambda _{tr}$ are usually believed to be
close to the superconducting ones. (A complete report of our calculated
transport properties will be given in the following subsection). For $V,$
the value of $\lambda _{tr}$ found by us is $1.15$ and, for $Nb,$ $\lambda
_{tr}=1.17.$ The calculated electrical and thermal resistivities are also
close to those measured. It therefore seems that our electron--phonon
coupling is accurate (within the computational accuracy of order $10\%)$
while the effect of the electron--electron interactions is small in these
metals.

As the next two examples, we report the results of our applications for $Mo$
and $Cu.$ There are no tunneling data for these materials because of their
low $T_c$ and the weakness of phonon effects. The calculated spectral
functions are presented in Fig.2(f),(g). Both curves qualitatively agree
with the corresponding phonon state densities shown for these metals in
Fig.1(f),(g) but a considerable frequency dependence of the electron--phonon
prefactor $\alpha ^2(\omega )$ (indicated by dashed lines) is also
predicted. For $Mo$ our linear--response calculations are found to be close
to our RMTA calculations and to earlier ones\cite{MoRMTA}. The estimated
average coupling here is $0.42$ which can be compared with the value $0.45$
deduced from the specific--heat measurements (see Table II). The calculated
value of $\lambda $ for $Cu$ is $0.14.$ The specific--heat estimate here is
less reliable possibly because of the smallness of $\lambda $ and errors due
to the experimental uncertainty in the value of $\gamma .$ Quite likely,
however, that there are some errors in the DFT value of the density of
states connected with the many-body effects since copper valence shell $%
3d^{10}$ is close to the strongly correlated $3d^9$ configuration. It is
known that within DFT the position of $d-$band is higher than experimentally
observed. The band which crosses the Fermi level is essentially $s-$band but
the effect of hybridization with the $d-$band should lead to lowering the
Fermi velocities. The latter effect is stronger if the $d-$band is closer to
the Fermi energy. To obtain $1+\lambda _{s-h}\sim 1.1$, one has to reduce
our calculated $N(\epsilon _F)$ by approximately $20\%$ which is a
reasonable estimate for the expected influence of the Coulomb correlations.
Concerning other estimates of $\lambda $ based on the transport properties,
our calculated $\lambda _{tr}=0.13$ while this value extracted from the
measured resistivity data is $0.12$ (see following subsection)$.$ Both
values are in agreement with our superconducting $\lambda .$

As the last example, we consider $Pd$ and discuss paramagnon effects. The
superconductivity in $Pd$ is absent because of the large spin fluctuations%
\cite{Parks}. There was also a discussion in the literature\cite{Rietschel}
on the paramagnon contributions to the mass enhancement in $Nb$ and $V$. The
occurrence of paramagnons is connected with the fluctuations of the electron
spins. Paramagnons usually counteract superconductivity since the latter has
its origin in the formation of pairs with the opposite spins. To extract the
paramagnon contribution, we can use our calculated values of $\lambda $
together with the specific--heat\cite{SH} estimates $\lambda _{s-h}$ after
formula (\ref{S3}). The necessary data are listed in Table II. Comparing
these results does not leave any place for $\lambda _{spin}=\lambda
_{s-h}-\lambda $ in all the materials except in $Pd$ which is a typical
example for paramagnon effects. Here $\lambda _{s-h}=0.69$ and with the use
of our calculated $\alpha ^2F(\omega )$ [Fig.2(h)], the average
electron--phonon coupling is found to be equal to $0.35.$ This results in
our value of $\lambda _{spin}=0.34$ for $Pd$ which is close to its earlier
estimate $0.31$ based on the RMTA\ calculation\cite{PinskiPD}.

After comparing the calculated and experimental spectral functions, we
present the results of our applications to solving the Eliashberg gap
equation with our knowledge of $\alpha ^2F(\omega ).$ Having fixed the
Coulomb pseudopotential $\mu ^{*}$, the superconducting state is now
completely described by the strong--coupling theory of superconductivity\cite
{Eliashberg}. According to the Allen-Dynes\cite{AllenDynes} modified McMillan%
\cite{McMillan} formula: 
\begin{equation}
T_c^{McM}=\frac{\omega _{log}}{1.2}\ exp\left( -\frac{1.04(1+\lambda )}{%
\lambda -\mu ^{*}(1+0.62\lambda )}\right)  \label{S1}
\end{equation}
the effect of the first reciprocal moment $\lambda $ of $\alpha ^2F(\omega )$
on $T_c$ is most important. Unfortunately, the estimation of the coupling
constant from $T_c$ is difficult because of the unknown value of $\mu ^{*}$.
We use standard Matsubara technique to solve numerically the Eliashberg
equation for $T_c$ and have found $\mu ^{*}$ which gives the experimental
value of $T_c\,$. The cutoff parameters $\omega _{cut}$ were taken to be
equal to ten phonon--boundary frequencies $\omega _{max}$. To treat the
Coulomb pseudopotential in terms of the expression (\ref{S1}) when solving
the Eliashberg equation we have rescaled actually used parameters $\mu
^{*}(\omega _{cut})$ to $\mu ^{*}=\mu ^{*}(\omega _{log})$ according to

\begin{equation}
\frac 1{\mu ^{*}}=\frac 1{\mu ^{*}(\omega _{cut})}+ln\frac{\omega _{cut}}{%
\omega _{log}}.  \label{S2}
\end{equation}
Table III reports the obtained $\mu ^{*}$ values. The main conclusion here
is that the calculated $\mu ^{*}$ varies between $0.11$ and $0.17$ which is
close to the conventional value usually taken $\sim 0.13.$ The noticeable
exceptions are only $Nb$ and, especially, $V$ for which too large $\mu ^{*}$
have been found. Of course, this overestimation occurs over the conventional
quantity $0.13$ while the detailed theoretical data on the Coulomb
pseudopotential are unknown. We think that the obtained quantities are still
below the upper limit for the allowed $\mu ^{*}$ values.

Also listed in Table III are the values of $T_c^{McM}$ evaluated after (\ref
{S1}) with our calculated $\omega _{log},$ $\lambda $ and $\mu ^{*}$. As it
can be seen, the exact solution of the Eliashberg equation gives $T_c$
(chosen to be the experimental value) which slightly deviates from that
estimated after the McMillan expression. The accuracy of the later is
averagely about $15\%$.

To conclude that our spectral functions provide a proper description of
superconductivity we have found the energy--gap parameters $\Delta _0$.
These results are shown in the last two rows of Table III. The available
tunneling data for $Al$ give $\Delta _0$ coinciding with the theoretical
value. Some overestimation of the coupling constant in comparing to the
experimental one has taken place in $Pb$, while $\Delta _0$ agrees very
closely. Let us turn out to the important case of transition metals $V,Nb$
and $Ta.$ As we have discussed already, the main difficulties in the
tunneling studies in $V$ and $Nb$ are connected with the oxidation of the
surface layers and the tunneling estimations of the coupling constant in $Nb$
vary considerably in the past. In contrast to it, the measured
superconducting gap in $Nb$ has approximately the same values in all the
experiments and is equal to $1.56$ $meV$\cite{Wolf}. This value perfectly
agrees with that found by us which is equal to $1.53$ $meV.$ We have also
found a good agreement in the energy gap for $V$ which is within $4\%$ of
the experimental one. The discrepancy in the energy gap for $Ta$ is again $%
1-2\%$. Because of the low transition temperature there is no tunneling data
for $Mo$ and we only give the theoretical value of $\Delta _0$. We thus see
that despite of the discrepancy in evaluating the coupling constants, an
extremely good agreement ($1-2\%$) is obtained for the predicted gap data.
Such a coincidence is readily understood because the\ ratio $2\Delta _0/T_c\,
$ is slowly varying for different superconductors. (It is $3.52$ within the
BCS\ theory.) Fixing the $T_c$ to its experimental value makes the value of $%
\Delta _0$ insensitive to the errors in $\alpha ^2F(\omega )$ and $\lambda .$

To summarize, we have found that our results for the spectral functions and,
in particular, for the coupling constants are realistic for the correct
description of the superconducting properties. Especially, an excellent
agreement has been found between our calculated $\lambda $ and the values
extracted from specific--heat measurements. The values of $\lambda $ deduced
from the available tunnelling experiments also agree within $10\%$ with our
calculations in all the materials except $Nb$ and, especially, $V.$ However,
taking into account the past tendency to correct the tunneling spectra for $%
Nb$ as well as our calculations for $Ta,$ we do not consider these
discrepancies as essential.

\smallskip\ 

{\bf d. Transport properties.}

\smallskip\ 

We now report the results of our applications for calculating the
electron--phonon contribution to the electrical and thermal resistivities
(conductivities). This field retains very important and interesting, first,
because the easily measured transport properties and, especially, the
electrical resistivities, provide a valuable information on the
electron--phonon--coupling strength and, second, no large--scale
investigations of these properties by ab initio theoretical calculations
appeared so far .

We calculate the electrical and thermal resistivity using the low--order
variational approximation (LOVA) and our theoretical transport spectral
functions found after Eq. (\ref{E12}). As follows from Eqs. (\ref{E13}) and (%
\ref{E14}) at high temperatures:

\begin{equation}
\rho =\frac{\pi \Omega _{cell}k_BT}{N(\epsilon _F)\langle v_x^2\rangle }%
\lambda _{tr},  \label{T1}
\end{equation}
\begin{equation}
w=\frac{6\Omega _{cell}}{\pi k_BN(\epsilon _F)\langle v_x^2\rangle }\lambda
_{tr},  \label{T2}
\end{equation}
and an important information is contained in the transport constant $\lambda
_{tr}$ defined by

\begin{equation}
\lambda _{tr}=2\int\limits_0^\infty \frac{d\omega }\omega \alpha
_{tr}^2F(\omega ).  \label{T3}
\end{equation}
It is usually believed that the latter is close to the superconducting $%
\lambda $ because the expressions for $\alpha _{tr}^2F(\omega )$ and $\alpha
^2F(\omega )$ are quite similar, except for the factor $[1-{\bf v}({\bf k})%
{\bf v}({\bf k}^{\prime })/|{\bf v}({\bf k})|^2]$ which preferentially
weights the backscattering processes. However, there may exist a significant
difference between $\alpha _{tr}^2F(\omega )$ and $\alpha ^2F(\omega )$ for
the case of strongly nested Fermi surfaces\cite{Grespi} due to the
contribution from the backscattering of electrons between the opposite sides
of the nested Fermi surface.

Despite the complexity of the Fermi surfaces in the transition metals we
have obtained the transport functions $\alpha _{tr}^2F(\omega )$ quite close
to the superconducting $\alpha ^2F(\omega )$ which have been shown in Fig.1.
The latter is also true for the simple metals considered in this work.
Unfortunately, we have not investigated an interesting question about the
low--frequency behavior of the $\alpha _{tr}^2F(\omega )$ due to a
relatively coarse grid of the phonon wave vectors used for integrating the
Eq. (\ref{E12}). The values of $\lambda _{tr}$ calculated from our transport
functions are listed in Table IV. Comparison between $\lambda _{tr}$ and
superconducting constants calculated earlier (Table II) gives the difference
between them within $20\%$ in all the tested materials. This is in agreement
with previous conclusions that $\lambda _{tr}\sim \lambda $ for transition
metals\cite{Pinski}.

The results of our calculated electrical resistivity $\rho (T)$ and thermal
conductivity $w^{-1}(T)$ are presented, respectively, in Fig.3 and Fig.4
(full lines), up to the temperatures $500K.$ Symbols denote different
measured points available from Refs. \onlinecite{LandoltTR1,LandoltTR2}.
(The residual values of the electrical resistivities are subtracted.) For
the comparison with the experiment we are limited by the temperatures $%
T<2\Theta _{tr},$ where $\Theta _{tr}\sim \sqrt{\left\langle \omega
^2\right\rangle _{tr}}$ is close to the average phonon energy. (We list our
calculated values of $\Theta _{tr}$ in Table IV.) This is so because the
description of the transport properties at high temperatures require to take
into account the unharmonicity effects and the Fermi--surface smearing. At
low temperatures (usually when $T<\Theta _{tr}/5$) the calculations will
demand the inclusion of the $N$--sheet and the inelasticity corrections
beyond LOVA\cite{Pinski}. Also, here a more careful integration over the
Brillouin zone is necessary to produce a correct limit of $\alpha
_{tr}^2F(\omega )$ when $\omega \rightarrow 0.$ Moreover, we cannot consider
very low temperatures because of the effects of electron--electron
scattering, size effects, impurity scattering, {\it etc,} which may give
considerable contributions in addition to the electron--phonon scattering.
The latter is basically responsible for the electrical resistivity of a
metal at high temperatures in the absence of spin fluctuations. For the
thermal conductivity, the lattice contribution to the heat current also
exist and must be taken into account at the temperatures at least less than $%
100K$. From Table IV we see that, except lead, the values of $\Theta _{tr}$
in all other materials are well above the low--temperature region and the
comparison of our results with the measured ones must be relevant at the
intermediate temperatures.

To compare the theoretical transport constants with the experiment, we fit
the measured data\cite{LandoltTR1} for $\rho (T)$ by polynomial series

\begin{equation}
\rho (T)=\sum\limits_{i=1}^nc_iT^{3-2i}  \label{T4}
\end{equation}
at the temperatures $\Theta _{tr}/2<T<2\Theta _{tr}$ with $n=2$. (The
accuracy of the fit varies within $3\%$ if $n$ is increased). The empirical
values $\lambda _{tr}^{exp}$ were then found using the extracted coefficient 
$c_1$ as follows

\begin{equation}
\lambda _{tr}^{exp}=\frac{c_1\omega _p^2}{8\pi ^2k_B}  \label{T5}
\end{equation}
where $\omega _p$ is our calculated bare plasma frequency

\begin{equation}
\omega _p^2=\frac{8\pi N(\epsilon _F)\langle v_x^2\rangle }{\Omega _{cell}}
\label{T6}
\end{equation}
The obtained $\lambda _{tr}^{exp}$ are shown in Table IV where they can be
compared with our calculated $\lambda _{tr}.$ Note that similar numerical
estimates for $\lambda _{tr}^{exp}$ can also be made by analyzing thermal
conductivities because the measured Lorentz number approaches to the
Sommerfeld value at the temperatures $T\geq \Theta _{tr}$.

For $Al$ [Fig.3(a)] we have found a good agreement between the theoretical
and the experimental\cite{LandoltTR1} resistivities at the whole interval of
the intermediate temperatures. The corresponding values of the transport
constants are $\lambda _{tr}^{calc}=0.37$ and $\lambda _{tr}^{exp}=0.39.$
The reduction of the coupling constant ($\lambda _{calc}=0.44$) to the
transport one is less than $20\%.$ There is also an agreement between the
theoretical curve and the experimental points\cite{LandoltTR2} for the
thermal conductivity [Fig.4(a)] above $150K$. The theory, however,
underestimates $w^{-1}(T)$ at the lower temperatures. An obvious explanation
here is the neglection of the lattice contribution to the thermal current.
In fact, as it can be seen from Fig.4, such underestimation at the low
temperatures exist in all other materials considered in this work.

A comparison with the experiment is complicated for lead because of its low
phonon energies ($\Theta _{tr}\sim 75K$) and the importance of the
unharmonicity effects already at the low temperatures. The latter can
possibly explain our discrepancy in the calculated electrical resistivity
behavior shown in Fig.3(b). The same disagreement exist in our results for
the thermal conductivity, Fig.4(b). Here, the discrepancy is minimal at the
temperatures near $75K$ and grows fastly as the temperature increases mainly
because of the linear decay of the measured thermal conductivity. This
obviously contradicts with the LOVA behavior of $w^{-1}(T)$ and is
consistent with our assumption on the importance of the unharmonicity.
Unlike in the other considered metals, the computed value $1.19$ of the
transport constant here is significantly smaller than the electron-phonon $%
\lambda =1.68.$ This reduction could also point out the importance of the
anisotropy in the electron-phonon scattering as well as the Fermi-surface
effects which are not well reproduced by LOVA. In the absence of a
calculation beyond LOVA it is difficult to determine the main source of
errors.

Measured resistivity for $Nb$ starts to saturate at high temperatures. In
fact, this effect is evident [Fig.3(d)] at the temperatures above $2\Theta
_{tr}\sim 400K$ and it does not appear at the intermediate interval where
the behavior of the resistivity only slightly deviates from the LOVA
prediction. Comparing the calculated $\lambda _{tr}=1.17$ and the empirical
values $\lambda _{tr}^{exp}=1.11$ gives the agreement about $5\%$. Like in $%
Nb,$ there is a complete agreement between the theoretical and the
experimental data for $V$ [Fig.3(c)] at the temperatures $\Theta
_{tr}/5<T<2\Theta _{tr}$ (calculated $\Theta _{tr}\sim 260K$ ). The
theoretical $\lambda _{tr}=1.15$ coincides with the $\lambda _{tr}^{exp}$
found empirically. The applicability of our method to the description of the
transport properties for both $V$ and $Nb$ is also supported by comparing
the thermal--conductivity data, Fig.4(c),(d).

Fig.3(e) and Fig.4(e) present the results of our calculations in $Ta.$ For $%
w^{-1}(T)$, an excellent correspondence of the theoretical prediction with
observed behavior is obtained, but $\rho (T)$ is underestimated in our
calculation within $10-12\%$. As a consequence, the evaluated $\lambda
_{tr}^{exp}=0.93$ slightly exceeds the value $0.83$ of the theoretical
transport constant. The discrepancy is, in principle, within our
computational errors. Note however, that the experimental behavior\cite
{LandoltTR1} of $\rho (T)$ used in evaluating $\lambda _{tr}^{exp}$ may
depend on the sample purity and should be verified by several measurements.
The disagreement can also be assigned to the influence of high--temperature
effects in the vicinity of the upper limit ($2\Theta _{tr}\sim 340K$) of the
intermediate interval .

The results for $Mo$ are given in Fig.3(f) and Fig.4(f). To avoid the
influence of high--temperature effects we have dropped out the measured
resistivity points at the temperatures above $300K$ (the calculated $\Theta
_{tr}\sim 290K)$ . Fitting for $T<300K$ gives the result for the empirical
transport constant $\lambda _{tr}^{exp}=0.40$ which is close to our
prediction, $\lambda _{tr}^{calc}=0.35$. The agreement in the
thermal--conductivity data is satisfactory for the whole intermediate
interval.

The slope of the resistivity in $Cu,$ Fig.3(g), is also obtained quite
accurate as in the other materials. The value of $\lambda _{tr}$ is only
slightly overestimated in the calculation. The discrepancy in the calculated
thermal conductivity, Fig.4(g), is larger and consists about $20\%$ at the
room temperature. We cannot explain such disagreement by the renormalization
due to the Coulomb correlations because, having been proportional to the
ratio $\lambda _{tr}/\omega _p^2$, both $\rho (T)$ and $w^{-1}(T)$ must be
insensitive to this effect in the first order. The underestimation of the
theoretical $w^{-1}(T)$ can point out the largeness of the lattice
contribution to the thermal conductivity at the temperatures above $100K.$
Unfortunately, there is a number of known difficulties to extract the latter
values from the experiments\cite{Williams}. From the low--temperature data
analysis\cite{Williams} one may conclude that the contribution to the
thermal resistivity, $w_{e-h}^p,$ from the process of a phonon decay by
emissing electron--hole pairs is very small for $Cu$ because of the
apparently weak electron--phonon coupling ($\lambda \sim 0.12-0.14)$. This
supports our explanation for the obtained discrepancy.

Finally, the predicted transport properties of $Pd$ are presented in
Fig.3(h) and Fig.4(h). Like $Cu,$ this metal has $4d^{10}$ electronic
configuration, but, in contrast to $Cu$, we have found a very good agreement
between the calculated curve $w^{-1}(T)$ and its measured behavior. We can
consequently judge that the thermal conductivity carried by phonons is small
in this case which is consistent with the conclusion\cite{Williams} that the
contribution $w_{e-h}^p$ is large for $Pd$. We have also found an
underestimation of the electrical resistivity in the calculation. The
agreement between $\lambda _{tr}$ and $\lambda _{tr}^{exp}$ (see Table IV)
consists about $15\%$ which is, in principle, the upper limit of our
computational uncertainty. Most likely, however, that the additional spin
fluctuation mechanism of the resistivity is also present in this metal.

In summary, the behavior of $\rho (T)$ and $w(T)$ is consistent with the
results (\ref{T1}),(\ref{T2}) at the intermediate temperatures and there is
no significant discrepancy between our calculations and the experimental
points. More precisely, we have extracted the values of $\lambda _{tr}^{exp}$
using the experimental data for $\rho (T)$ together with our band--structure
value of $\omega _p$ and found the agreement between the experimental and
the theoretical transport constants to be about $10\%$ (in particular, lower
than $5\%$ for $Al,Nb,Mo$ and $V$). In fact, compared with the experiment is
the ratio $\lambda _{tr}/\omega _p^2.$ Except possibly $Cu$ and $Pd,$ the
DFT--based band--structure calculations assumed to provide the proper
magnitude for the plasma frequency. So, we drop the possibility of error
cancellations and conclude that the theoretical $\lambda _{tr}$ are in the
real agreement with the experiment. Relatively high error level in $Pb$ can
be explained by the importance of the unharmonic effects. The large lattice
contribution to the thermal conductivity could affect our comparison for $Cu.
$ In $Pd,$ the additional mechanism of the resistivity can also take place.
Nevertheless, taking into account the agreement in the other calculated
properties, we think that our description of the electronic transport in the
considered materials is quite satisfactory.

\smallskip\ 

{\bf IV. CONCLUSION}

\smallskip\ 

We have presented {\it ab initio} linear-response calculations of the
electron-phonon interaction in the transition metals $Cu,Mo,Nb,Pd,Ta,V$ and
in the $sp$-metals $Al,Pb$ using the local density functional method and the
LMTO basis set. Our results for the lattice dynamical, superconducting and
transport properties in these materials agree well with the experiment. They
can be summarized as follows: (i) we have obtained tunneling spectral
functions $\alpha ^2F(\omega )$ and their first reciprocal moments $\lambda $
close to the measured ones; (ii) the correct values for the superconducting
energy gap have been found using our calculated $\alpha ^2F(\omega )$ and $%
\mu ^{*}$ corresponding to the experimental $T_c;$ (iii) the solution of the
Eliashberg equation for $T_c$ (or for $\mu ^{*}$ if $T_c$ is fixed) is well
approximated by the conventional McMillan formula; (iv) the mass enhancement
observed in the specific-heat measurements corresponds very well to our
calculations and there is no paramagnon contribution in all the metals
except $Pd$; (v) we have found the electrical and thermal resistivities in
agreement with the measured data; (vi) we have also found them to be well
described by the LOVA\ expressions; (vii) the theoretical transport
constants agree with the values of $\lambda _{tr}^{exp}$within $10\%$. To
summarize all these results, we conclude that our method gives the
description of the electron-phonon coupling with the accuracy of order $10\%.
$ We also conclude that the effect of renormalization of the energy bands
due to electron--electron interactions is small in the considered materials.
Some discrepancies between the theoretical and the tunneling values of $%
\lambda $ in $Nb$ and $V$ can be assigned to the difficulty in processing
the tunneling data. Nevertheless, it seems to us that more experimental and
theoretical work is necessary to account for the large $\lambda $ and $\mu
^{*}$ in $Nb$ and, especially, in $V.$

\smallskip\ 

{\bf Acknowledgments.}

The authors are indebted to O. K. Andersen, E. G. Maksimov, O. V. Dolgov, 
O. Jepsen, A. Liechtenstein, I. I. Mazin, and S. Shulga for many helpful
discussions. One of us (D. Y. S) was partially supported by INTAS(93-2154),
ISF(MF-8300) and RFFI grants.

\bigskip\

%TCIMACRO{\TeXButton{endtwocolumn}{\end{multicols}}}
%BeginExpansion
\end{multicols}
%EndExpansion

%%%%%%%%%%%%%%%%%%%
%%%%%%%%%%%%%%%%%%%
%  TABLES
%%%%%%%%%%%%%%%%%%%
%%%%%%%%%%%%%%%%%%%

\bigskip\ 

%TCIMACRO{\TeXButton{B}{\begin{table}[tbp] \centering}}
%BeginExpansion
\begin{table}[tbp] \centering
%EndExpansion
\caption{Comparison between calculated and experimental\protect\cite{Landolt} phonon 
frequencies (THz) at the high--symmetry points $X,L$ for the fcc metals $Al,Pb,Cu,Pd$, and at the points $H,N$ for the 
bcc metals $V,Nb,Ta,Mo$. Also listed are the theoretical--to--experimental volume ratios 
$V/V_0$ used in the calculations.} 
\begin{tabular}{llllllllll}
fcc$\Vert $bcc &  & $Al$ & $Pb$ & $V$ & $Nb$ & $Ta$ & $Mo$ & $Cu$ & $Pd$ \\ 
\hline
$X_L\Vert H_{LT}$ & 
\begin{tabular}{l}
theory \\ 
exp.
\end{tabular}
& 
\begin{tabular}{l}
$9.51$ \\ 
$9.69$%
\end{tabular}
& 
\begin{tabular}{l}
$1.80$ \\ 
$1.86$%
\end{tabular}
& 
\begin{tabular}{l}
$8.03$ \\ 
$-$%
\end{tabular}
& 
\begin{tabular}{l}
$6.43$ \\ 
$6.49$%
\end{tabular}
& 
\begin{tabular}{l}
$5.13$ \\ 
$5.03$%
\end{tabular}
& 
\begin{tabular}{l}
$5.71$ \\ 
$5.52$%
\end{tabular}
& 
\begin{tabular}{l}
$7.69$ \\ 
$7.25$%
\end{tabular}
& 
\begin{tabular}{l}
$7.17$ \\ 
$6.72$%
\end{tabular}
\\ 
&  &  &  &  &  &  &  &  &  \\ 
$X_T\Vert N_L$ & 
\begin{tabular}{l}
theory \\ 
exp.
\end{tabular}
& 
\begin{tabular}{l}
$5.83$ \\ 
$5.78$%
\end{tabular}
& 
\begin{tabular}{l}
$1.06$ \\ 
$0.89$%
\end{tabular}
& 
\begin{tabular}{l}
$7.22$ \\ 
$-$%
\end{tabular}
& 
\begin{tabular}{l}
$5.52$ \\ 
$5.66$%
\end{tabular}
& 
\begin{tabular}{l}
$4.54$ \\ 
$4.35$%
\end{tabular}
& 
\begin{tabular}{l}
$7.99$ \\ 
$8.14$%
\end{tabular}
& 
\begin{tabular}{l}
$5.36$ \\ 
$5.13$%
\end{tabular}
& 
\begin{tabular}{l}
$5.01$ \\ 
$4.64$%
\end{tabular}
\\ 
&  &  &  &  &  &  &  &  &  \\ 
$L_L\Vert N_{T_1}$ & 
\begin{tabular}{l}
theory \\ 
exp.
\end{tabular}
& 
\begin{tabular}{l}
$9.84$ \\ 
$9.69$%
\end{tabular}
& 
\begin{tabular}{l}
$2.18$ \\ 
$2.18$%
\end{tabular}
& 
\begin{tabular}{l}
$4.76$ \\ 
$-$%
\end{tabular}
& 
\begin{tabular}{l}
$3.94$ \\ 
$3.93$%
\end{tabular}
& 
\begin{tabular}{l}
$2.65$ \\ 
$2.63$%
\end{tabular}
& 
\begin{tabular}{l}
$5.74$ \\ 
$5.73$%
\end{tabular}
& 
\begin{tabular}{l}
$7.77$ \\ 
$7.30$%
\end{tabular}
& 
\begin{tabular}{l}
$7.39$ \\ 
$7.02$%
\end{tabular}
\\ 
&  &  &  &  &  &  &  &  &  \\ 
$L_T\Vert N_{T_2}$ & 
\begin{tabular}{l}
theory \\ 
exp.
\end{tabular}
& 
\begin{tabular}{l}
$4.33$ \\ 
$4.19$%
\end{tabular}
& 
\begin{tabular}{l}
$0.92$ \\ 
$0.89$%
\end{tabular}
& 
\begin{tabular}{l}
$6.17$ \\ 
$-$%
\end{tabular}
& 
\begin{tabular}{l}
$4.80$ \\ 
$5.07$%
\end{tabular}
& 
\begin{tabular}{l}
$4.18$ \\ 
$4.35$%
\end{tabular}
& 
\begin{tabular}{l}
$4.69$ \\ 
$4.56$%
\end{tabular}
& 
\begin{tabular}{l}
$3.64$ \\ 
$3.42$%
\end{tabular}
& 
\begin{tabular}{l}
$3.60$ \\ 
$3.34$%
\end{tabular}
\\ 
&  &  &  &  &  &  &  &  &  \\ 
$V/V_0$ &  & $0.955$ & $1.002$ & $0.990$ & $0.972$ & $0.974$ & $0.971$ & $%
0.985$ & $0.975$%
\end{tabular}
%TCIMACRO{\TeXButton{E}{\end{table}}}
%BeginExpansion
\end{table}
%EndExpansion

\bigskip\ 

%TCIMACRO{\TeXButton{B}{\begin{table}[tbp] \centering}}
%BeginExpansion
\begin{table}[tbp] \centering
%EndExpansion
\caption{Comparison between the calculated electron--phonon 
coupling constants $\lambda _{calc}$
and the values of $\lambda _{tun}$ deduced from the tunnelling 
experiments. 
Also listed are the values of $\lambda _{s-h}$ extracted from the 
measured specific--heat coefficient $\gamma$ with
the use of our calculated density of states $N(\epsilon _F)$.} 
\begin{tabular}{lllllllll}
& $Al$ & $Pb$ & $V$ & $Nb$ & $Ta$ & $Mo$ & $Cu$ & $Pd$ \\ \hline
$\lambda _{calc}$ & $0.44$ & $1.68$ & $1.19$ & $1.26$ & $0.86$ & $0.42$ & $%
0.14$ & $0.35$ \\ 
$\lambda _{tun}$ & $0.42^a$ & $1.55^a$ & $0.82^a$ & $1.04^a,1.22^b$ & $%
0.78^a $ & $-$ & $-$ & $-$ \\ 
$\lambda _{s-h}$ & $0.43$ & $1.64$ & $1.00,1.17$ & $1.17$ & $0.83$ & $0.45$
& $-$ & $0.69$ \\ 
&  &  &  &  &  &  &  &  \\ 
$N(\epsilon _F),\frac{states}{Ry*cell}$ & $5.49$ & $6.87$ & $26.14$ & $20.42$
& $18.38$ & $8.34$ & $4.36$ & $34.14$ \\ 
$\gamma ,\frac{mJ}{K^2mol}$ & $1.36^c$ & $3.14^c$ & $9.04^c,9.82^d$ & $%
7.66^c $ & $5.84^c$ & $2.10^c$ & $0.69^c$ & $10.0^c$%
\end{tabular}
%TCIMACRO{
%\TeXButton{E}{$^a$Reference \protect\onlinecite{Wolf} \\
%$^b$Reference \protect\onlinecite{Bostock1} \\
%$^c$Reference \protect\onlinecite{SH} \\
%$^d$Reference \protect\onlinecite{Vedeneev}
%\end{table}}}
%BeginExpansion
$^a$Reference \protect\onlinecite{Wolf} \\
$^b$Reference \protect\onlinecite{Bostock1} \\
$^c$Reference \protect\onlinecite{SH} \\
$^d$Reference \protect\onlinecite{Vedeneev}
\end{table}
%EndExpansion

\bigskip\ 

%TCIMACRO{\TeXButton{B}{\begin{table}[tbp] \centering}}
%BeginExpansion
\begin{table}[tbp] \centering
%EndExpansion
\caption{Calculated values of the Coulomb pseudopotential $\mu ^{*}$ which 
provide the experimental values of $T_c$ as the solutions of the Eliashberg
equation with our knowledge of $\alpha ^2 F(\omega)$.
Values $T_c^{McM}$ were then found with our $\omega _{log}$, 
$\lambda $ and $\mu ^{*}$
in order to check the accuracy of the McMillan $T_c$ --expression . 
Also shown are the computed and the measured\protect\cite{Wolf} superconducting 
energy-gap parameters $\Delta _0$.} 
\begin{tabular}{lllllllll}
& $Al$ & $Pb$ & $V$ & $Nb$ & $Ta$ & $Mo$ & $Cu$ & $Pd$ \\ \hline
$\mu ^{*}$ & $0.12$ & $0.17$ & $0.30$ & $0.21$ & $0.17$ & $0.14$ & $0.11$ & $%
-$ \\ 
&  &  &  &  &  &  &  &  \\ 
$T_c^{exp},K$ & $1.18$ & $7.19$ & $5.40$ & $9.25$ & $4.47$ & $0.92$ & $>0$ & 
$-$ \\ 
$T_c^{McM},K$ & $1.22$ & $6.81$ & $6.68$ & $10.5$ & $5.11$ & $0.67$ & $>0$ & 
$-$ \\ 
&  &  &  &  &  &  &  &  \\ 
$\omega _{log},K$ & $270$ & $65$ & $245$ & $185$ & $160$ & $280$ & $220$ & $%
180$ \\ 
&  &  &  &  &  &  &  &  \\ 
$\Delta _0^{calc},meV$ & $0.18$ & $1.35$ & $0.84$ & $1.53$ & $0.70$ & $0.14$
& $-$ & $-$ \\ 
$\Delta _0^{exp},meV$ & $0.18$ & $1.33$ & $0.81$ & $1.56$ & $0.71$ & $-$ & $%
- $ & $-$%
\end{tabular}
%TCIMACRO{\TeXButton{E}{\end{table}}}
%BeginExpansion
\end{table}
%EndExpansion

\bigskip\ 

%TCIMACRO{\TeXButton{B}{\begin{table}[tbp] \centering}}
%BeginExpansion
\begin{table}[tbp] \centering
%EndExpansion
\caption{Comparison between calculated and empirical values of the transport
constant $\lambda _{tr}$. The values of $\lambda _{tr}^{exp}$ 
were deduced from the electrical--resistivity data\protect\cite{LandoltTR1}
with help of our calculated bare plasma frequencies $\omega _p$.
Also listed are the average transport frequencies 
$\Theta _{tr}=\protect\sqrt{\left\langle \omega ^2\right\rangle _{tr}}$} 
\begin{tabular}{lllllllll}
& $Al$ & $Pb$ & $V$ & $Nb$ & $Ta$ & $Mo$ & $Cu$ & $Pd$ \\ \hline
$\lambda _{tr}^{calc}$ & $0.37$ & $1.19$ & $1.15$ & $1.17$ & $0.83$ & $0.35$
& $0.13$ & $0.43$ \\ 
$\lambda _{tr}^{exp}$ & $0.39$ & $1.52$ & $1.15$ & $1.11$ & $0.93$ & $0.40$
& $0.12$ & $0.50$ \\ 
&  &  &  &  &  &  &  &  \\ 
$\omega _p,eV$ & $12.29$ & $14.93$ & $7.95$ & $9.47$ & $9.05$ & $8.81$ & $%
8.75$ & $7.34$ \\ 
&  &  &  &  &  &  &  &  \\ 
$\Theta _{tr},K$ & $330$ & $75$ & $260$ & $200$ & $170$ & $290$ & $230$ & $%
190$%
\end{tabular}
%TCIMACRO{\TeXButton{E}{\end{table}}}
%BeginExpansion
\end{table}
%EndExpansion

%%%%%%%%%%%%%%%%%%%%%%%%%%
%%%%%%%%%%%%%%%%%%%%%%%%%%
%%%%% FIGURE CAPTIONS.
%%%%%%%%%%%%%%%%%%%%%%%%%%
%%%%%%%%%%%%%%%%%%%%%%%%%%

\bigskip\ 

Fig.1(a)-(h). Calculated phonon--dispersion curves along several symmetry
directions for the eight elemental metals considered in this work. The lines
result from the interpolation between the theoretical points (circles) . The
results of avaiable neutron--difraction measurements\cite{Landolt} are shown
by triangles. Also plotted are the calculated densities of states (DOS).

\bigskip\ 

Fig.2(a)-(h). Calculated spectral functions $\alpha ^2F(\omega )$ of the
electron--phonon interaction (full lines) for the eight elemental metals
considered in this work. The behavior of the electron--phonon prefactor $%
\alpha ^2(\omega )$ is shown by dashed lines. Symbol plots present the
results of available tunnelling experiments\cite{Wolf,Lead}.

\bigskip\ 

Fig.3(a)-(h). Calculated temperature dependence of the electrical
resistivity, $\rho (T),$ as a lowest--order variational solution of the
Boltzmann equation for the eight elemental metals considered in this work.
Symbols show different experimental data available from Ref. %
\onlinecite{LandoltTR1}.

\bigskip\ 

Fig. 4(a)-(h). Calculated temperature dependence of the thermal
conductivity, $w^{-1}(T),$ as a lowest--order variational solution of the
Boltzmann equation for the eight elemental metals considered in this work.
Symbols show different experimental data available from Ref. %
\onlinecite{LandoltTR2}.

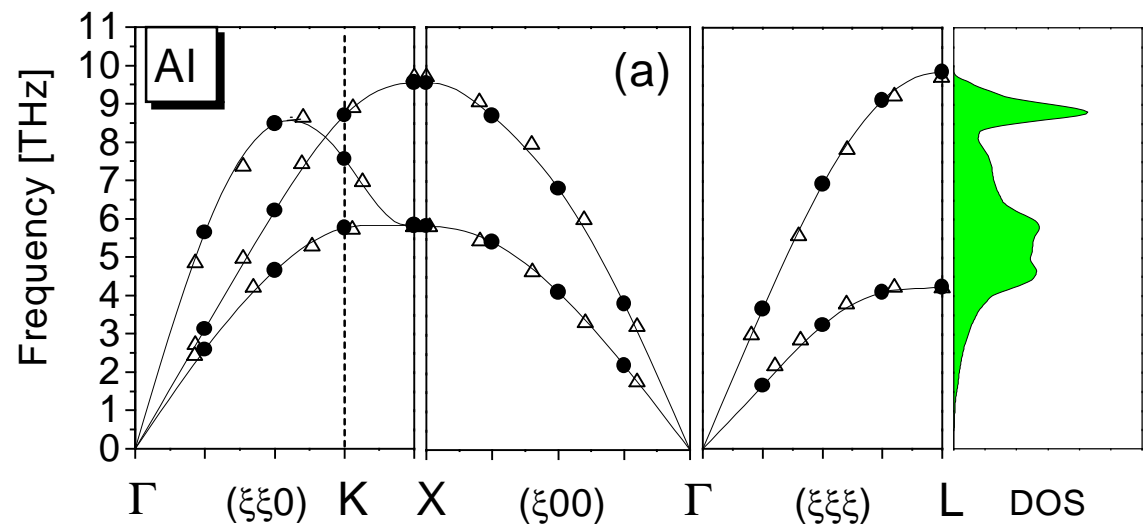
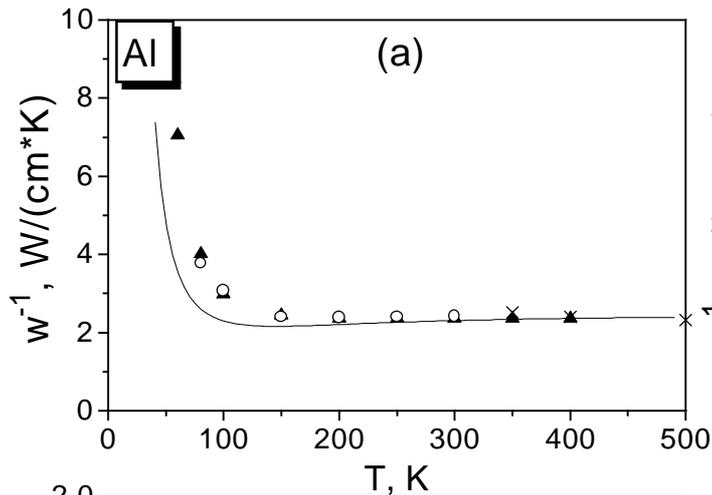
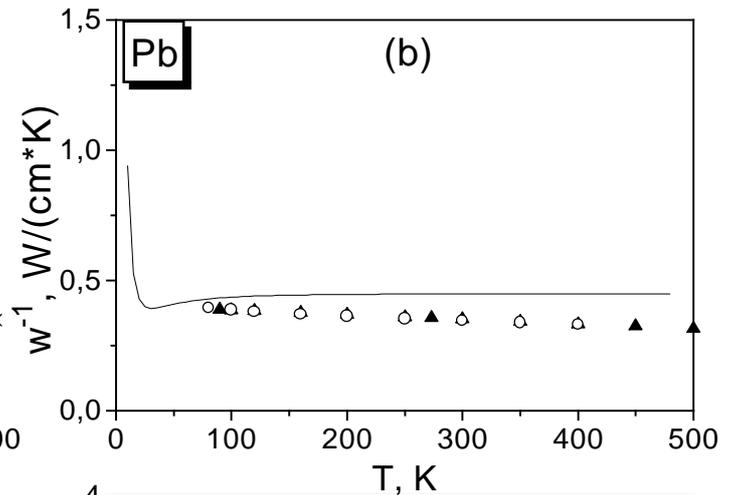
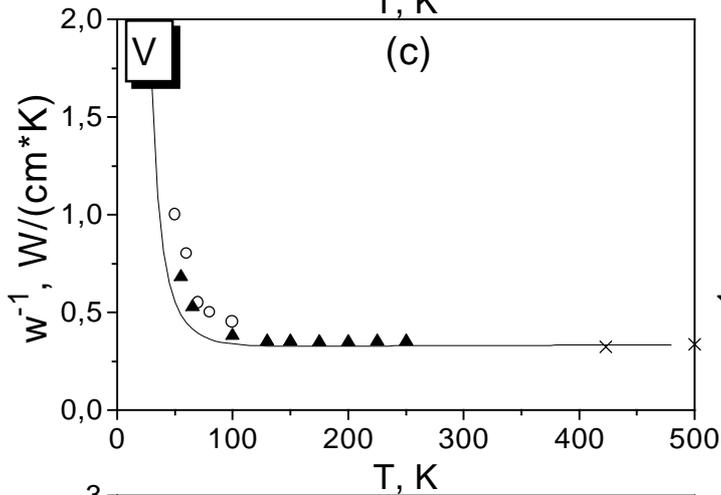
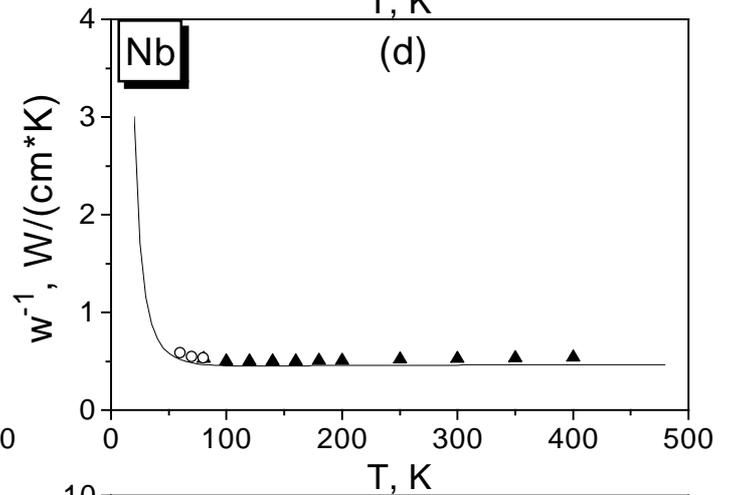
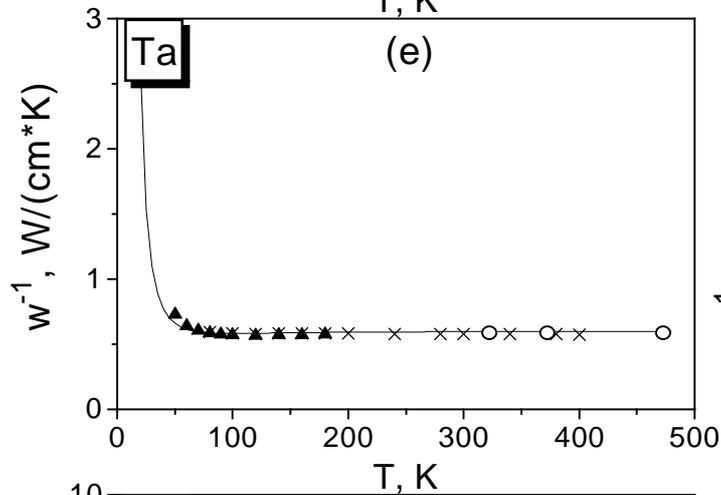
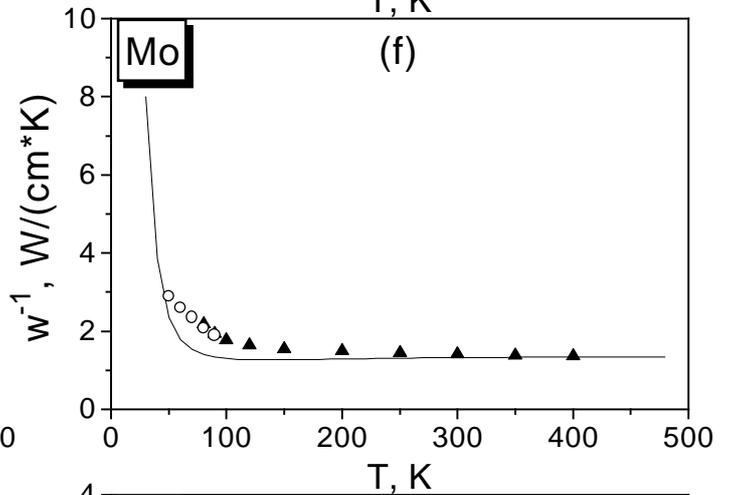
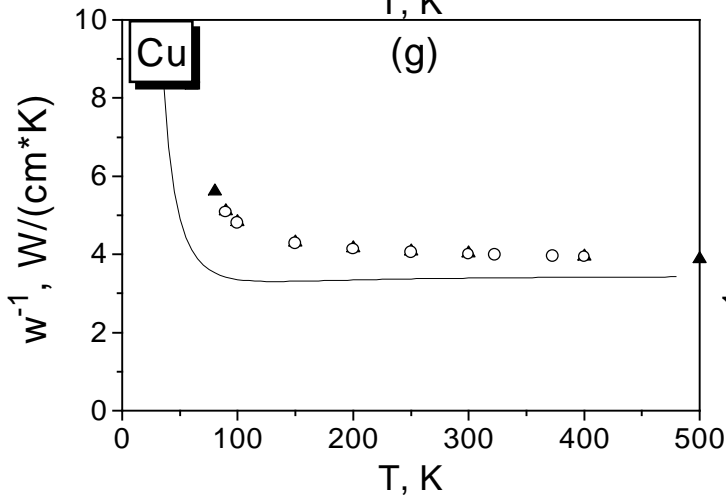
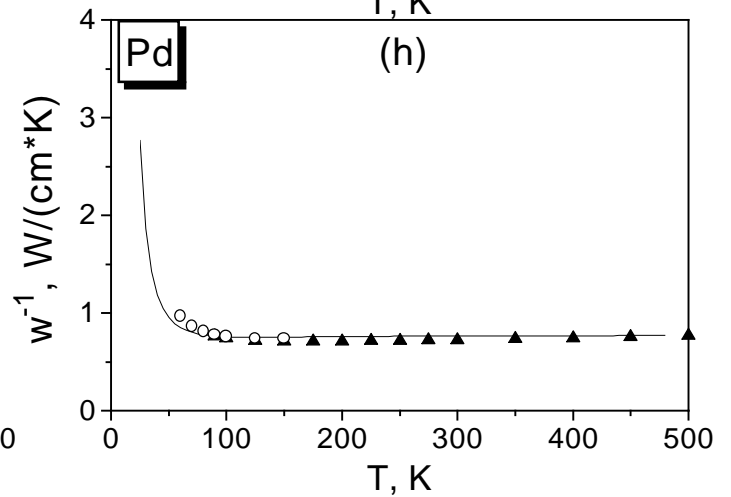
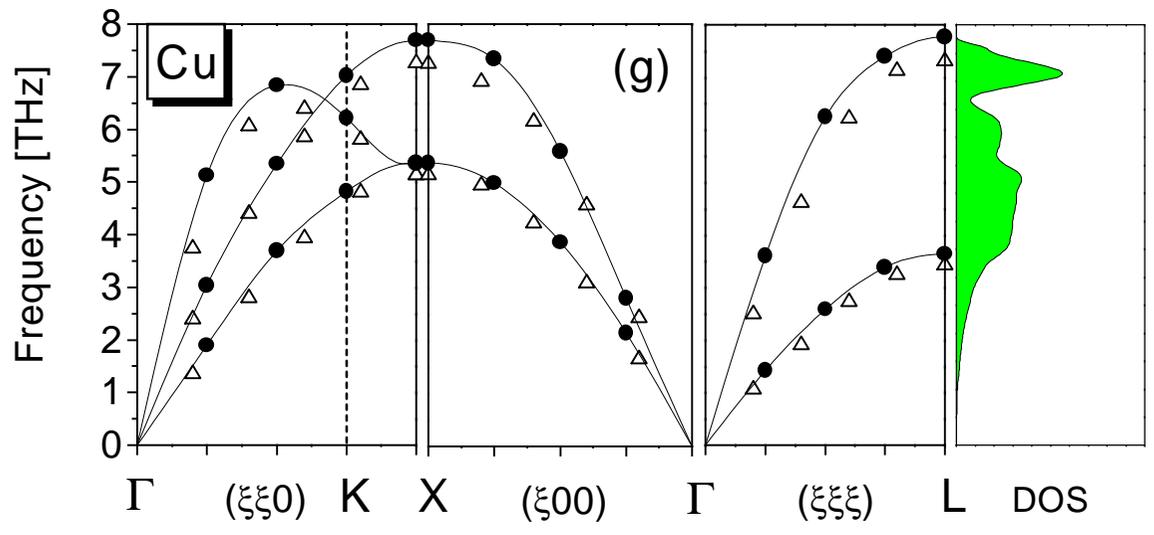
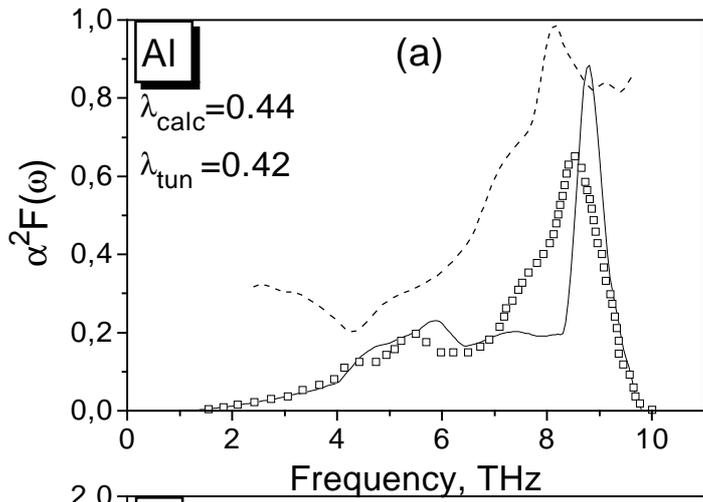
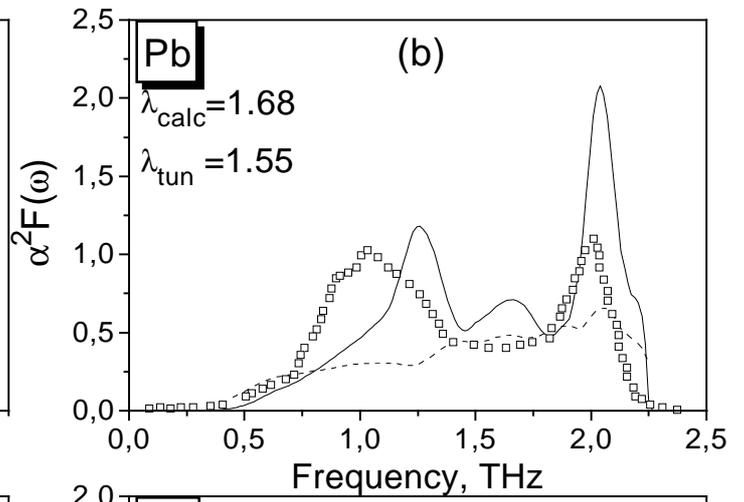
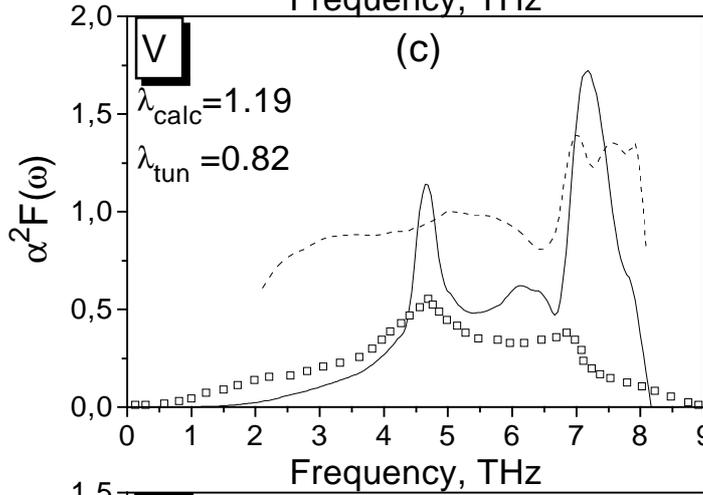
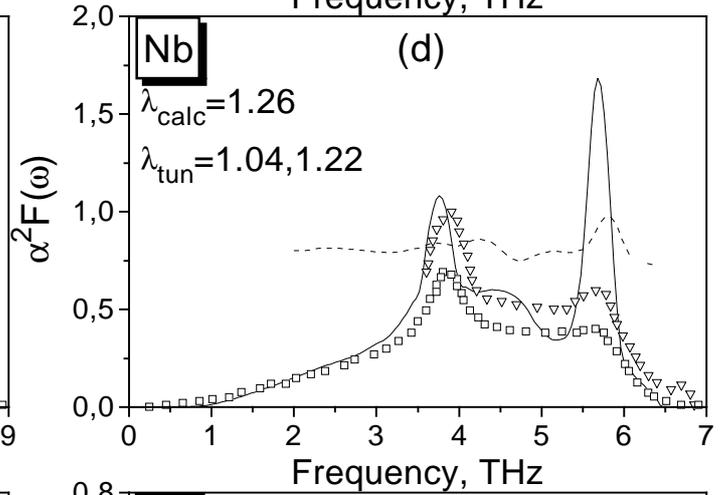
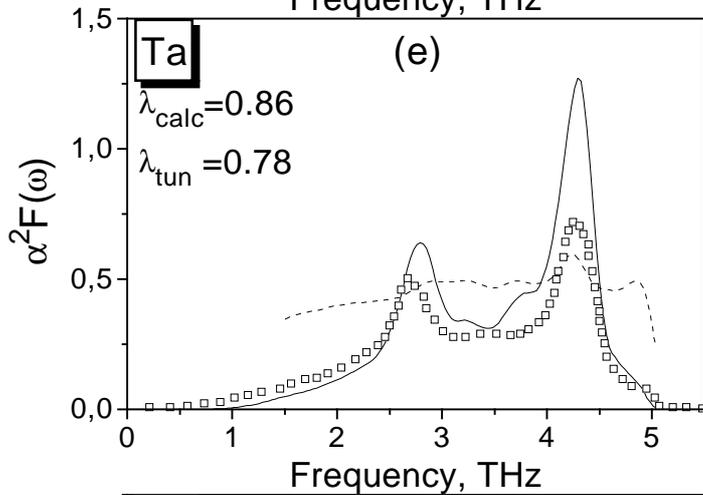
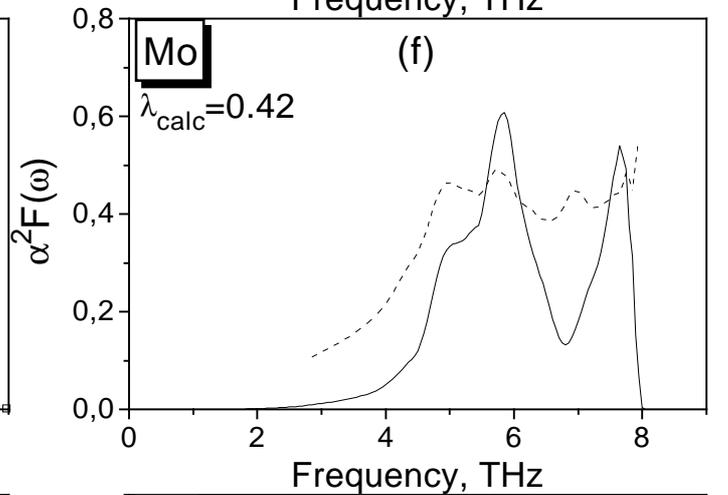
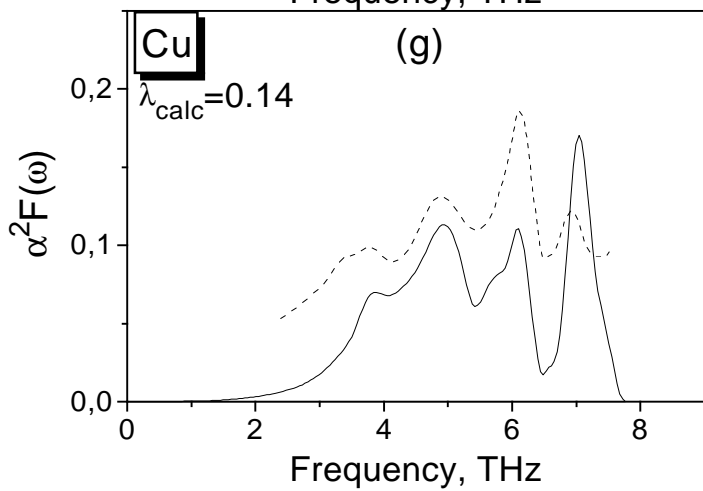
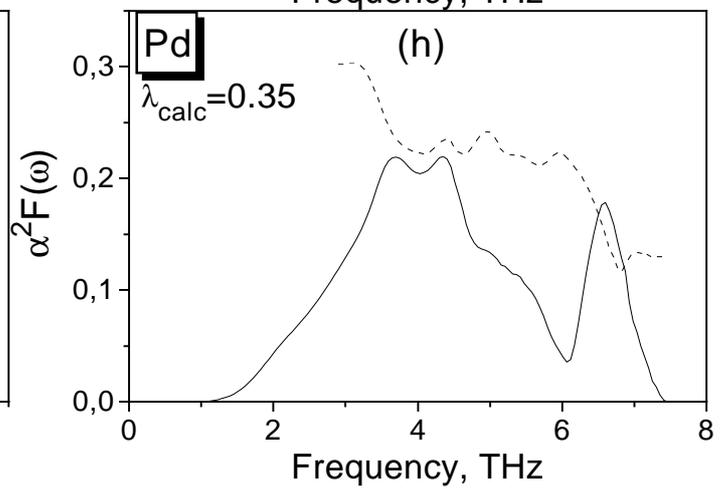
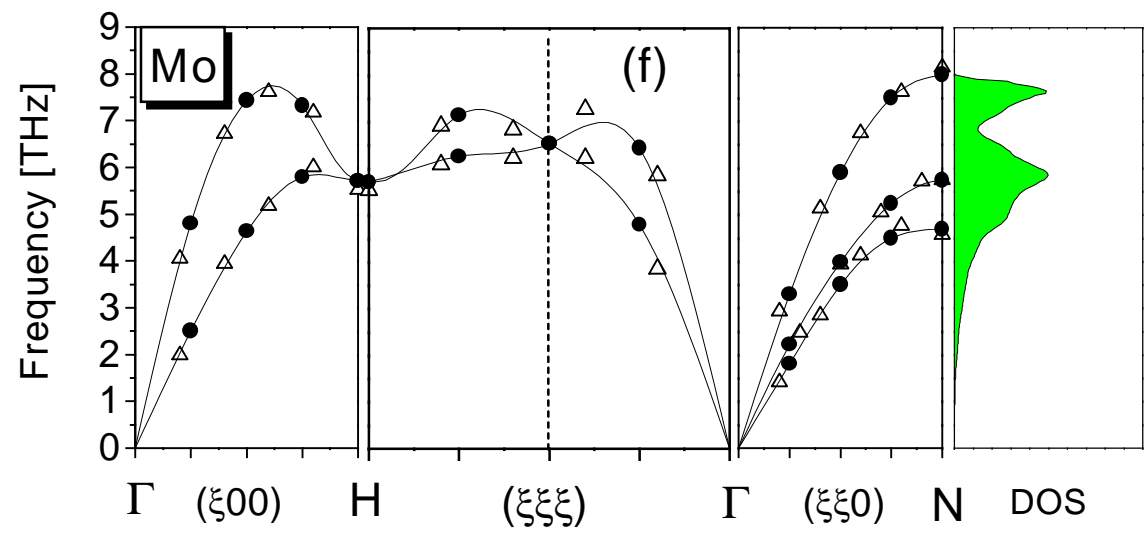
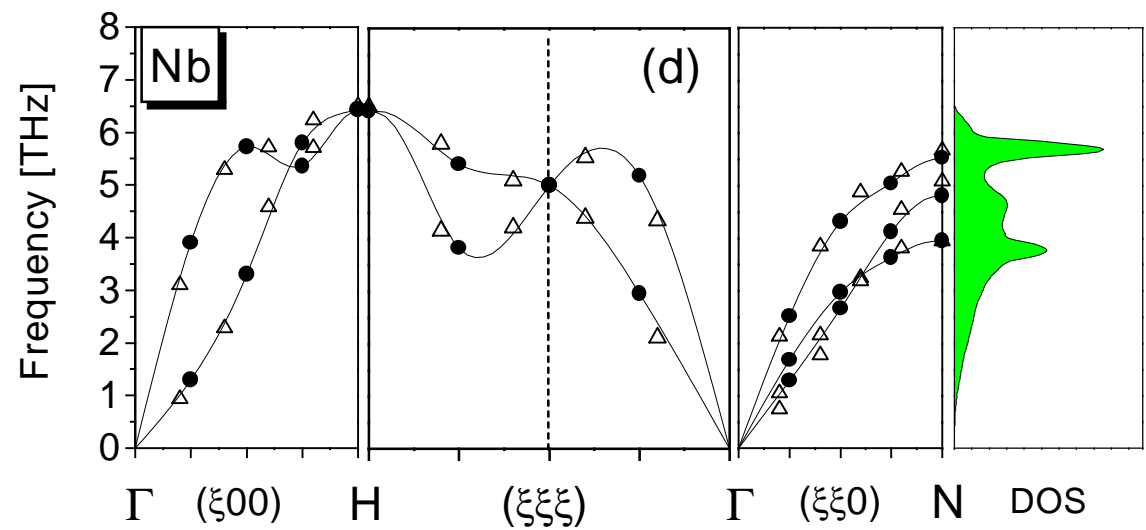
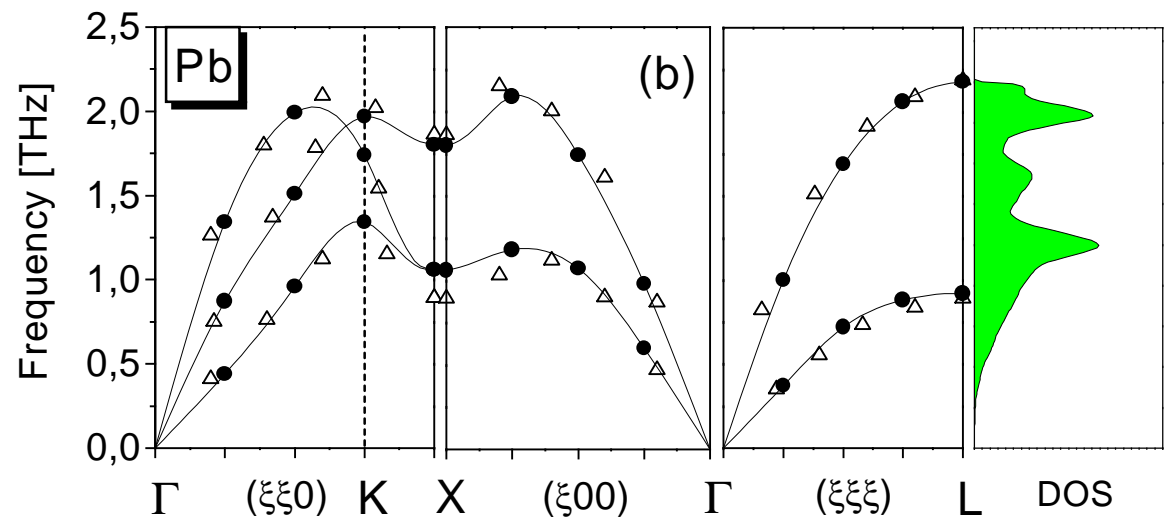
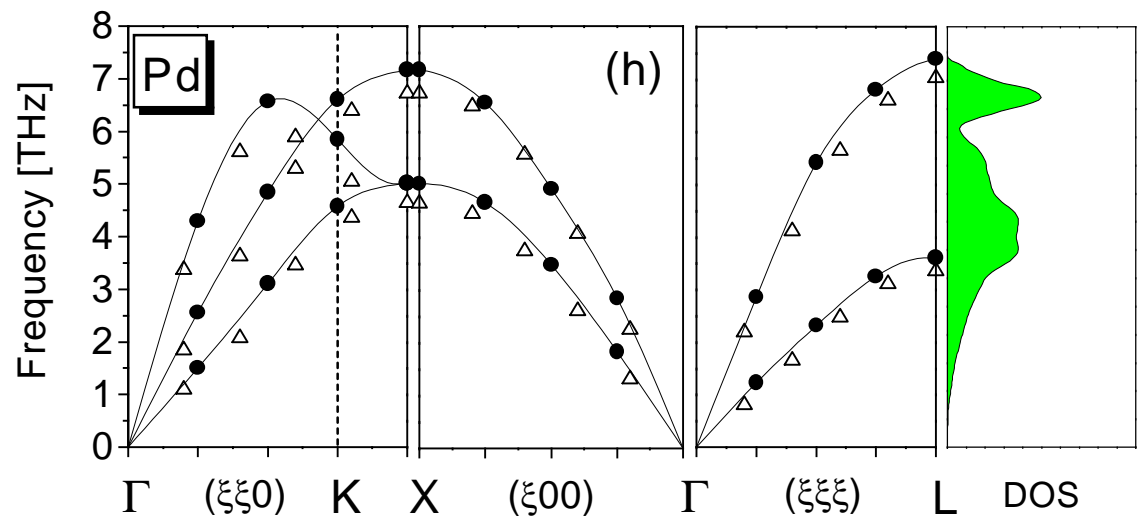
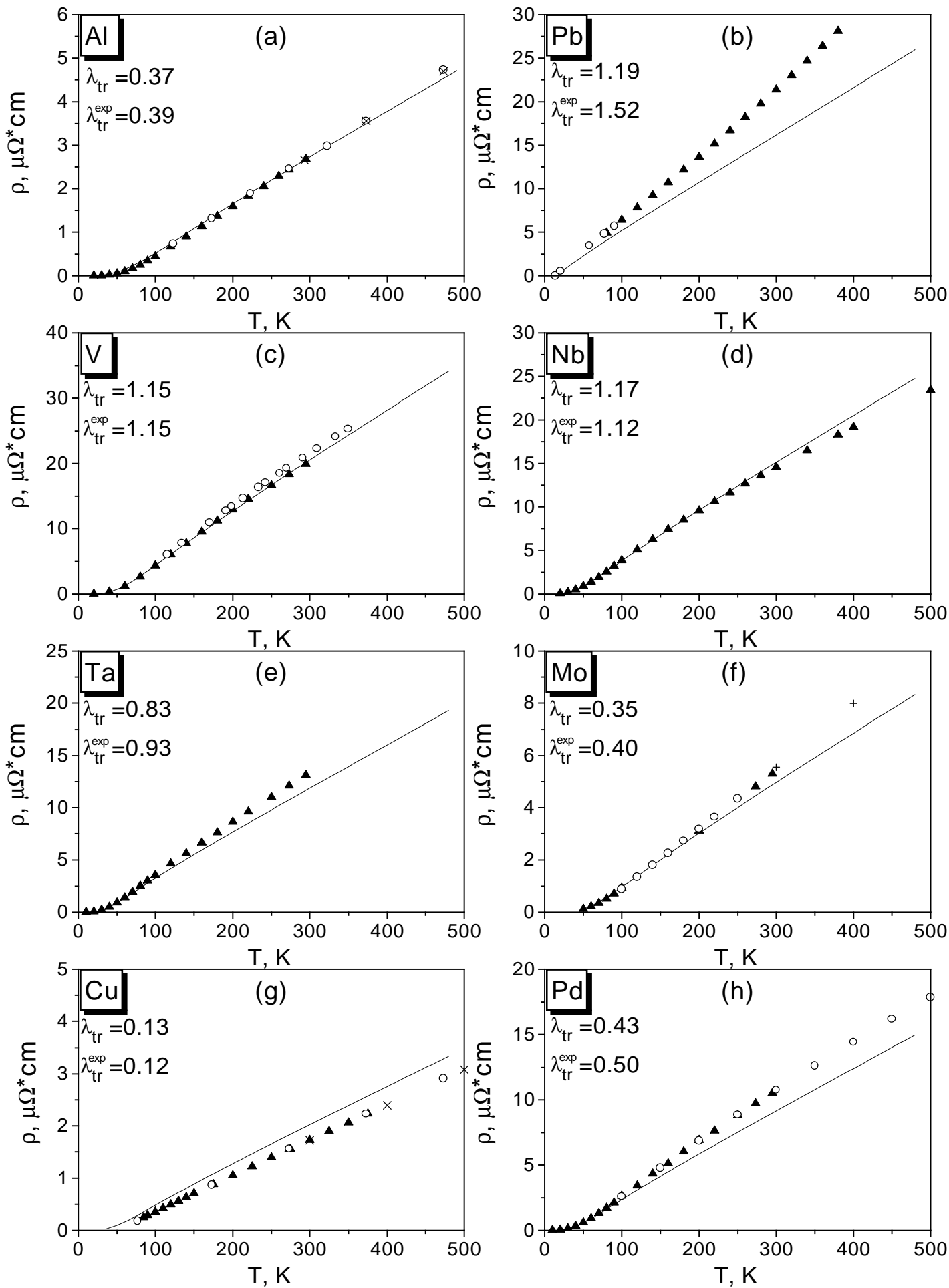
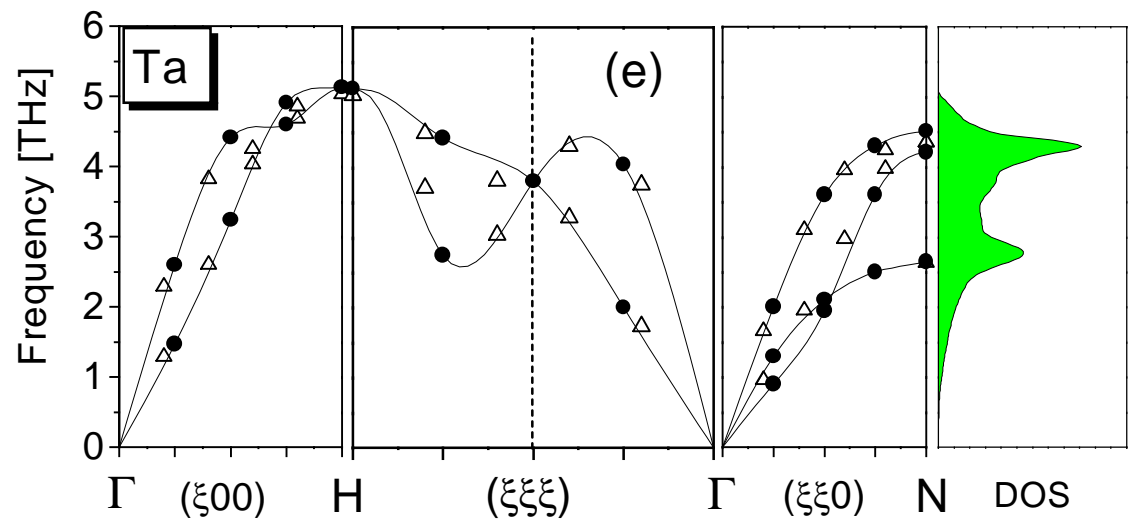
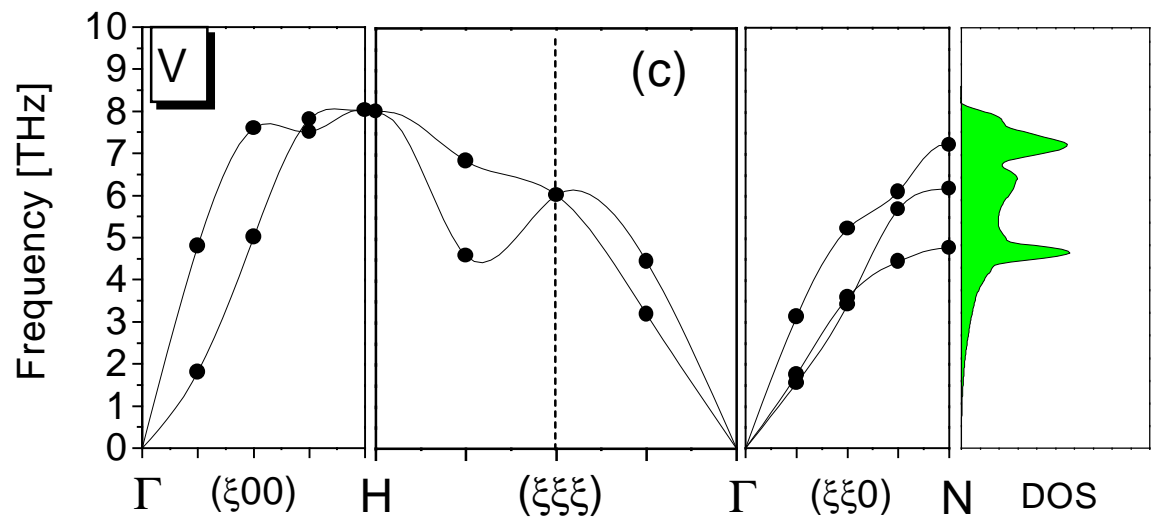
\end{document}